 \documentclass[12pt,preprint]{aastex6}
\usepackage{natbib}
\newcommand{\qsonumber}{48}
\newcommand{\andromedadistance}{752}
\newcommand{\beamsize}{\ensuremath{9\farcm1}}

\newcommand{\virialCovering}{0.051}
\newcommand{\MSmetallicity}{\ensuremath{-0.95\pm0.12}}
\newcommand{\HI}{\ensuremath{\mbox{\ion{H}{1}}}}

\newcommand{\OI}{\ensuremath{\mbox{\ion{O}{1}}}}
\newcommand{\SiII}{\ensuremath{\mbox{\ion{Si}{2}}}}
\newcommand{\SiIII}{\ensuremath{\mbox{\ion{Si}{3}}}}
\newcommand{\CII}{\ensuremath{\mbox{\ion{C}{2}}}}
\newcommand{\MgII}{\ensuremath{\mbox{\ion{Mg}{2}}}}
\newcommand{\lya}{Ly$\alpha$\relax}
\newcommand{\lyb}{Ly$\beta$\relax}

\newcommand{\NHI}{\ensuremath{N(\mbox{\ion{H}{1}})}\relax}
\newcommand{\logNHI}{\ensuremath{\log N(\mbox{\ion{H}{1}})}\relax}

\newcommand{\Rvir}{\ensuremath{R_{\rm vir}}\relax}

\newcommand{\msun}{\ensuremath{M_\odot}\relax}

\newcommand{\column}{cm$^{-2}$}
\newcommand{\percc}{cm$^{-3}$}

\newcommand{\kms}{km~s$^{-1}$}
\newcommand{\hst}{{\em HST}}

\newcommand{\vlsr}{\ensuremath{v_{\rm LSR}\relax}}

\newcommand{\fccum}{\ensuremath{f_c(\le \rho)}\relax}

\newcommand{\lstar}{\ensuremath{L^*}}
\newcommand{\coshalos}{COS-Halos}

\defcitealias{lehner2015}{LHW15}
\defcitealias{hafen2017}{H17}

\begin{document}

\title{Project AMIGA: A Minimal Covering Factor for Optically Thick
  Circumgalactic Gas Around the Andromeda Galaxy}

\author{
  J. Christopher Howk\altaffilmark{1,2},
  Christopher B. Wotta\altaffilmark{1},
  Michelle A. Berg\altaffilmark{1},
  Nicolas Lehner\altaffilmark{1},
  Felix J. Lockman\altaffilmark{3},
  Zachary Hafen\altaffilmark{4},
  D.J. Pisano\altaffilmark{5,6},
  Claude-Andr\'{e} Faucher-Gigu\`{e}re\altaffilmark{4},
  Bart P. Wakker\altaffilmark{7},
  J. Xavier Prochaska\altaffilmark{8,9}
  Spencer A. Wolfe\altaffilmark{5,6},
  Joseph Ribaudo\altaffilmark{10},
  Kathleen A. Barger\altaffilmark{11},
  Lauren Corlies\altaffilmark{12},
  Andrew J. Fox\altaffilmark{13},
  Puragra Guhathakurta\altaffilmark{8,9},
  Edward B. Jenkins\altaffilmark{14},
  Jason Kalirai\altaffilmark{12,13},
  John M. O'Meara\altaffilmark{15},
  Molly S. Peeples\altaffilmark{12,13},
  Kyle R. Stewart\altaffilmark{16}
  Jay Strader\altaffilmark{17}}

%%%%%%%%%%%%%%%%%%%%%%%%%

\altaffiltext{1}{Department of Physics, University of Notre Dame,
  Notre Dame, IN, USA}
\altaffiltext{2}{Visiting professor, Instituto de Astrof\'{i}sica,
  Pontificia Universidad Cat\'{o}lica de Chile, Santiago, Chile}
\altaffiltext{3}{Green Bank Observatory, Green Bank,
  WV, USA}
\altaffiltext{4}{Department of Physics and Astronomy and Center for
  Interdisciplinary Exploration and Research in Astrophysics (CIERA),
  Northwestern University, Evanston, IL, USA}
\altaffiltext{5}{Department of Physics \& Astronomy, West Virginia
  University, Morgantown, WV, USA}
\altaffiltext{6}{Center for Gravitational Waves and Cosmology, West
  Virginia University, Chestnut Ridge Research Building, Morgantown,
  WV, USA}
\altaffiltext{7}{Department of Astronomy, University of
  Wisconsin-Madison, Madison, WI, USA}
\altaffiltext{8}{Department of Astronomy and Astrophysics, University
  of California, Santa Cruz, CA, USA}
\altaffiltext{9}{University of California Observatories, Lick
  Observatory, Santa Cruz, CA, USA}
\altaffiltext{10}{Department of Physics, Utica College, Utica, NY, USA}
\altaffiltext{11}{Department of Physics \& Astronomy, Texas Christian
  University, Fort Worth, TX, USA}
\altaffiltext{12}{Department of Physics and Astronomy, Johns Hopkins University, Baltimore, MD, USA}
\altaffiltext{13}{Space Telescope Science Institute, Baltimore, MD, USA}
\altaffiltext{14}{Princeton University Observatory, Princeton, NJ, USA}
\altaffiltext{15}{Department of Chemistry and Physics, Saint Michael's
  College, Colchester, VT, USA}
\altaffiltext{16}{Department of Mathematical Sciences, California Baptist University, 8432 Magnolia Ave., Riverside, CA, USA}
\altaffiltext{17}{Department of Physics and Astronomy, Michigan State
  University, East Lansing, MI, USA}

%%%%%%%%%%%%%%%%%%%%%%%%%%%%%%%%%%%%%%%%%%%%%%%%%%%%%%%%%%%%%%%%%%%%%%

%\altaffiltext{1}{Based on data acquired ...}
%
%%%%%%%%%%%%%%%%%%%%%%%%%%%%%%%%%%%%%%%%%%%%%%%%%%%%%%%%%%%%%%%%%%%%%%

\begin{abstract}

  We present a deep search for \HI\ 21-cm emission from the gaseous
  halo of Messier 31 as part of Project AMIGA, a large Hubble Space
  Telescope program to study the circumgalactic medium of the
  Andromeda galaxy. Our observations with the Robert C. Byrd Green
  Bank Telesope target sight lines to \qsonumber\ background AGNs,
  more than half of which have been observed in the ultraviolet with
  the Cosmic Origins Spectrograph, with impact parameters
  $25 \la \rho \la 330$ kpc ($0.1 \la \rho / \Rvir \la 1.1$).  We do
  not detect any 21-cm emission toward these AGNs to limits of
  $\NHI \approx 4 \times10^{17}$ \column\ ($5\sigma$; per 2 kpc
  diameter beam). This column density corresponds to an optical depth
  of $\sim2.5$ at the Lyman limit; thus our observations overlap with
  absorption line studies of Lyman limit systems at higher
  redshift. Our non-detections place a limit on the covering factor of
  such optically-thick gas around M31 to $f_c < \virialCovering$ (at
  90\% confidence) for $\rho\le \Rvir$.  While individual clouds have
  previously been found in the region between M31 and M33, the
  covering factor of strongly optically-thick gas is quite small.  Our
  upper limits on the covering factor are consistent with expectations
  from recent cosmological ``zoom'' simulations. Recent \coshalos\
  ultraviolet measurements of \HI\ absorption about an ensemble of
  galaxies at $z \approx 0.2$ show significantly higher covering
  factors within $\rho \la 0.5 \Rvir$ at the same \NHI, although the
  metal ion-to-\HI\ ratios appear to be consistent with those seen in
  M31.

\end{abstract}

%%%%%%%%%%%%%%%%%%%%%%%%%%%%%%%%%%%%%%%%%%%%%%%%%%%%%%%%%%%%%%%%%%%%%%


\section{Introduction}

The gaseous circumgalactic medium (CGM) around galaxies serves as a
massive reservoir of baryons and metals \citep{peeples2014, werk2014,
  keeney2017} and an import driver of galactic evolution. For example,
gas and metals accreted from the CGM may provide the fuel for as much
as half of a massive galaxy's stellar mass by $z\sim0$
\citep[e.g.,][]{oppenheimer2010, ford2014, angles-alcazar2016}. The
CGM includes contributions from a broad range of sources. It includes
material ejected from the central galaxy in the form of winds
\citep[e.g.,][]{weiner2009, tripp2011, shen2013, rubin2014, cafg2015,
  muratov2017}, which may ultimately return to the galaxy. The CGM
includes matter removed from satellites as gas stripped via tidal
forces or ram pressure \citep{grcevich2009, spekkens2014, emerick2016}
or gas ejected via the winds of those same satellites
\citep[e.g.,][]{angles-alcazar2016}.\footnote{These are not
  necessarily mutually exclusive, as ram pressure may play a role in
  removing wind-driven matter from the satellite's halo.} It likely
also includes matter that has accreted from the IGM, either as cold
gas or gas that is heated via its interaction with the existing CGM
\citep{keres2005, fumagalli2011, van-de-voort2011,cafg2011}.  Most of
these processes produce (at least initially) relatively dense, cool
concentrations of gas. This is the case even for galactic winds, for
which a significant fraction of the mass ejected may come from cool,
entrained or condensed material \citep[e.g.,][]{heckman1990,
  schwartz2004, rupke2005, rubin2014, kacprzak2014}. These cool
streams of matter may ultimately fall onto the central galaxy itself,
perhaps fueling future star formation \citep{keres2005,
  oppenheimer2010, ford2014} or dissipate within the hotter, more
diffuse coronal matter in the halo \citep[e.g.,][]{joung2012,
  voit2015}. Thus, the observable CGM about galaxies is expected to be
multiphase, with the very diffuse warm and hot gas
\citep[$10^5 \la T \la 10^7$ K;][]{tumlinson2011, prochaska2011,
  wakker2009, anderson2013} threaded with denser cool gas
\citep[$10^4 \la T \la 10^5$ K;][]{nielsen2013, kacprzak2013,
  werk2014, keeney2017}, including the dense structures that could
represent new matter flowing into the CGM for the first time
\citep{lehner2013, wotta2016, van-de-voort2012}.

In the context of the CGM, such dense ($n_{\rm H} \ga 10^{-3}$ \percc)
and cool ($T < 10^5$ K) structures are ionized entities. They have
characteristic densities and size scales that produce optical depths
at the Lyman limit of order unity and above \citep{schaye2001,
  fumagalli2011a, cafg2011, van-de-voort2012}. In QSO absorption line
studies of the CGM, these streams would thus be categorized as Lyman
limit systems \citep[LLSs; ][]{tytler1982,steidel1990}, with \HI\
column densities $N(\mbox{\HI}) \ge 1.6\times10^{17}$ \column\
($\logNHI \ge 17.2$), which corresponds to an optical depth at the H
ionization edge of $\tau_{\rm 1 Ryd} \ge 1$
\citep{spitzer1978}. Studies of low-redshift Lyman limit systems have
shown they exhibit a broad range of metallicities, from $<1\%$ solar
to super-solar metallicities \citep{lehner2013, wotta2016}, as
expected if they are tracing a range of origins, including accreting,
stripped, or expelled matter.

Simulations of galaxies at $z\approx2$ show that optically-thick \HI\
material may cover $\sim2\% - 10\%$ of the area within the virial
radius of typical galaxies \citep[e.g.,][]{fumagalli2011a, cafg2011},
with a signicantly higher covering fraction at the smallest impact
parameters. Only recently have simulations been able to regularly
study the high-density structures in the CGM to $z=0$ with full
feedback prescriptions. The covering factors of optically thick gas in
simulated galaxies at low redshift tend to be lower on average than
their higher-redshift counterparts, partly due to the general
dimunition expected in cold accretion as well as generally diminished
feedback.  For example, \citealt{hafen2017} (hereafter
\citetalias{hafen2017}) find a median $f_c \sim 1\%$ for
$\log M_{\rm halo}/\msun \approx 12$ galaxies at $z=0$. Still, the
covering factors for individual galaxies can be as high as
$f_c \sim 10\% - 15\%$, depending on the feedback recipes and the
histories of recent star formation, mergers, and accretion
\citep[][\citetalias{hafen2017}]{gutcke2016}. Measurements of
optically-thick \HI\ thus constrain the total mass in the densest
material associated flows of matter through the CGM while also
constraining prescriptions for feedback and other processes in galaxy
simulations.

The study of galactic halos at $\logNHI \sim 17-18$ has largely been
limited to absorption line measurements. However, the the best radio
observatories now reach sensitivities to \HI\ 21-cm emission that get
nearly to $\tau_{\rm 1 \, Ryd} \approx 1$ gas around low-redshift
galaxies \citep[e.g.,][]{braun2004, lockman2012, wolfe2013,
  wolfe2016}, bridging the gap between emission and absorption
experiments. Radio searches for faint emission have the distinct
advantage of (potentially at least) providing more spatial information
than can be gleaned from the small numbers of skewers that are
available for absorption line spectroscopy, with its need for bright
background sources (especially in the ultraviolet).

Here we use the Robert C. Byrd Green Bank Telescope (GBT), part of the
Green Bank Observatory,\footnote{The Green Bank Observatory is a
  facility of the National Science foundation operated under a
  cooperative agreement by Associated Universities, Inc.} to search
for 21-cm emission from the halo of the nearby Andromeda galaxy. This
21-cm search is in support of our Project AMIGA (Absorption Maps In
the Gas of Andromeda)\footnote{Not to be confused with the AMIGA
  survey (Analysis of the interstellar Medium in Isolated GAlaxies, PI
  Lourdes Verdes-Montenegro); {\tt http://amiga.iaa.es}.}, a large
{\em Hubble Space Telescope} (\hst) program to study the absorption
from the extended CGM about the Andromeda galaxy (M31) using the
Cosmic Origins Spectrograph (COS). Project AMIGA will measure UV
absorption lines along $\sim$25 AGN sight lines passing within
$\rho \approx \Rvir \approx 300$ kpc of the center of M31. This builds
on our recent study demonstrating that M31 has a massive, diffuse CGM
stretching to approximately \Rvir\ \citep[][hereafter
\citetalias{lehner2015}]{lehner2015}. Project AMIGA will characterize
the radial and azimuthal dependences of metal ion surface densities
(for \ion{C}{2}, \ion{C}{4}, \ion{Si}{2} \ion{Si}{3}, \ion{Si}{4}, and
others) in M31's CGM. However, our \hst\ observations will provide
little information on the \HI\ distribution at the velocities where
\citetalias{lehner2015} find metal absorption from M31's CGM
($\vlsr \ga -500$ \kms), as the strongly-saturated \lya\ absorption
from the Milky Way completely blocks the light from the background
AGNs over these wavelengths. This is not a problem with \HI\ emission,
where the Milky Way emission is largely constrained to higher
velocities than M31's CGM.

In the present work we target \qsonumber\ directions through the
Andromeda galaxy's halo in order to measure the covering factor of
highly optically-thick gas ($\tau_{\rm 1\, Ryd} \approx 2.5$). Our
observations achieve sensitivities to \HI\ column densities that
overlap with absorption line measurements of CGM \HI\ about more
distant galaxies \citep[e.g.,][]{tumlinson2013, lehner2013,
  prochaska2017} and UV absorption line studies of high-velocity
clouds in the CGM of our own Milky Way \citep[e.g.,][]{fox2006}. As
the 21-cm line is so easy to excite (the critical density for
excitation is $n_{\rm crit} \approx 3\times10^{-5}$ \percc , well
below the density of gas we are considering, and Ly$\alpha$ pumping is
also an effective excitation mechanism), these GBT observations allow
us to trace gas even in conditions expected of the CGM.

There is good reason to suspect that M31 may have a significant CGM,
even one detectable in 21-cm emission. The relative dearth of neutral
gas associated with dwarf satellite galaxies has led several groups to
to argue for the existence of a low-density diffuse CGM about M31 to
impact parameters $\rho \la 270$ kpc \citep[][ see also
\citealt{spekkens2014} for similar work specific to the Milky
Way]{blitz2000, grcevich2009}. Similarly, although the giant southern
stream has a dynamical age of $<1$ Gyr \citep{fardal2008}, its star
formation was shut off well before it was tidally disrupted
\citep[$\ga 4$ Gyr ago;][]{brown2006}, suggesting its gas was stripped
by M31's CGM long ago. \citeauthor{grcevich2009} used ram pressure
stripping arguments to imply a density of a few $\times10^{-4}$
\percc\ for a diffuse (potentially hot) corona about M31.
\citet{rao2013} searched for metal line absorption from the halo of
M31, but their sensitivity was limited by the low spectral resolution
of their data. \citetalias{lehner2015} found significant metal ion
surface densities for sight lines projected within
$\rho \approx \Rvir \approx 300$ kpc of M31, suggesting an extended
distribution of low-density gas.  The total CGM gas mass implied by
the \citetalias{lehner2015} observations is significant at
$M_{\rm total} \ga 2\times10^8\, (Z/Z_\odot)^{-1}$ \msun\ within
$0.2\Rvir$, although the total mass could be a factor of 10 larger if
the covering factor is near unity to the virial radius.\footnote{The
  results initially given in \citetalias{lehner2015} are too high by a
  factor of $\sim10$ due to an unfortunate error.}

The \HI\ emission from the diffuse CGM implied by these studies may be
minimal, depending on the metallicity and ionization state of the
gas. In the Andromeda system, \citet{braun2004} reported the detection
of a diffuse ``bridge'' of low column density material stretching
between M31 and its companion M33 (projected $\sim200$ kpc from M31),
with an extension several degrees beyond M31 in the north
($1^\circ \approx 13$ kpc at M31). In total \citeauthor{braun2004}
found the filament stretched nearly 260 kpc.  This work was done with
a coarse angular resolution of $49\arcmin$ and velocity resolution of
18 \kms\ to achieve a sensitivity of $\logNHI \approx 17$ for emission
filling the beam, and much of the bridge emission appeared to have
beam-averaged column densities well below $\logNHI \sim
18$. Subsequent high-resolution maps of the vicinity of M33 have
called into questioned the presence of this bridge, given the lack of
a bridge component in the maps of M33
\citep{putman2009}. \citet{lockman2012} confirmed the existence of
faint 21-cm emission in this region between the two galaxies, but they
noted the gas must be very patchy, as they did not find it
consistently over several GBT pointings. \citet{wolfe2013} and
\citet{wolfe2016} showed that most of the emission identified by
\citet{braun2004} was associated with higher column density,
small-scale clumps of gas. The ``bridge'' appears to be constituted of
small clouds (each with masses $\sim10^5$ \msun) that appeared as more
continuous, diffuse structure when diluted within the Westerbork beam
employed by the \citeauthor{braun2004} study.  And while
\citet{wolfe2016} did detect some emission in the extended structure
found by \citet{braun2004} to the northwest of M31, the spatial
distribution of this gas appears to be much different than that
implied by the initial maps. Thus, M31's CGM appears to harbor at
least some small-scale concentrations of \HI\ detectable through their
21-cm emission.

In the inner regions of M31's halo, \citet{thilker2004} identified a
population of high-velocity clouds (HVCs) projected within
$\rho \sim 50$ kpc of the center of M31, which were subsequently
mapped at higher resolution by \citet{westmeier2005}. These clouds
individually have \HI\ masses $M_{\rm HI} \approx 10^5 - 10^6$ \msun\
with size scales of $\sim1$ kpc and central column densities
$\NHI \sim10^{19}$ \column\ \citep{westmeier2005}. They are discrete
structures, dense clouds ($n_{\rm HI} \approx 10^{-2}$ \percc) in the
CGM that may be similar to the clouds identified at larger distances
by \citet{wolfe2016}, though with higher masses.

In the present experiment (Project AMIGA GBT), we survey sight lines
toward UV-bright AGNs with impact parameters $\rho \la \Rvir$ from M31
in order to characterize the distribution of dense, cool gas in the
CGM. In no cases have the observed directions been chosen with prior
knowledge of the local \HI\ content. Thus, we can use our observations
to directly determine the covering factor of optically-thick gas in
the CGM of M31. Our covering factor estimates are complementary to the
\citet{richter2012} assessment of the covering factor of HVCs about
M31 based on the \citet{thilker2004} maps (which have a $5\sigma$
column density sensitivity of $\logNHI \sim18.25$ per GBT beam for a
FWHM$ = 25$ \kms, significantly higher than our
observations).\footnote{\citet{thilker2004} discuss several variations
  of their datacube, each smoothed to different velocity and spatial
  resolutions. Here we give the limits associated with their
  higher-resolution data, with a $\approx13\farcm65$ beam and 18 \kms\
  velocity channels. Their limits imply a mass sensitivity of
  $M_{\HI} \le 3.5\times 10^{4}$ \msun\ per beam.}

Our paper is presented as follows. We summarize the GBT observations
and data reduction in \S \ref{sec:observations}. We derive covering
factors for cool gas about the Andromeda galaxy in \S
\ref{sec:coveringfactor}. In \S \ref{sec:comparisons} we consider the
covering factor results in the context of recent QSO absorption line
measurements and numerical simulations. Our discussion and summary
follow in \S \ref{sec:discussion} and \S \ref{sec:summary}. A
metallicity measurement for a sight line passing through the nearby
Magellanic Stream (MS) is included in an Appendix.

Throughout we assume a distance of \andromedadistance\ kpc to M31
\citep{riess2012}. For comparison with other galaxies (simulated and
observed), we assume for M31 a stellar mass
$\log M_*/\msun \approx 11.0$ and a halo mass
$\log M_{\rm h}/\msun \approx 12.0$ \citep{tamm2012}.

%%%%%%%%%%%%%%%%%%%%%%%%%%%%%%%%%%%%%%%%%%%%%%%%%%%%%%%%%%%%%%%%%%%%%%
%%
%%

\section{GBT 21-cm Observations and Analysis}
\label{sec:observations}

\begin{figure}
  \epsscale{1.2}
\plotone{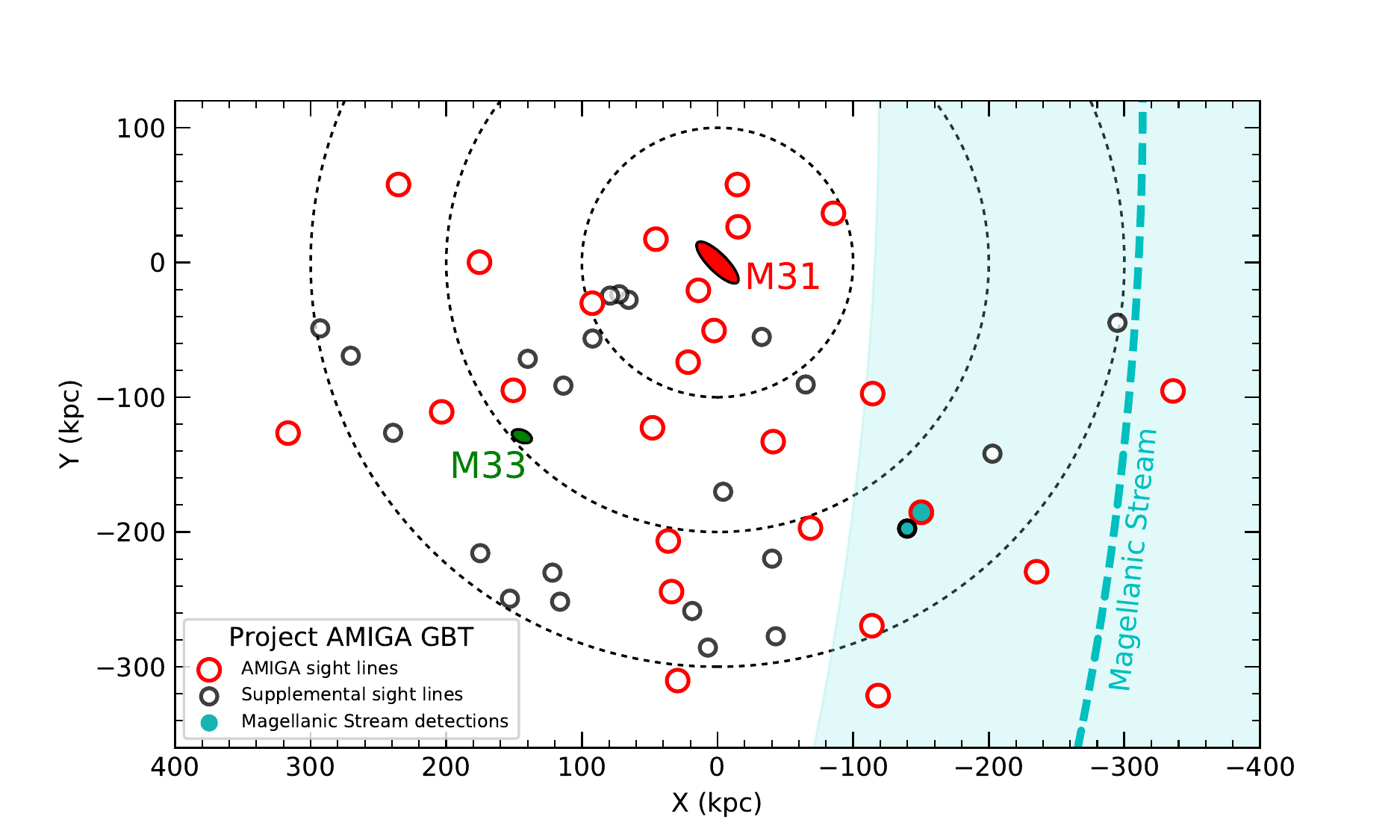}
% pyplot_mapgbt.py
\caption{The locations of our GBT pointings relative to the M31-M33
  system, where the axes show physical impact parameter from the
  center of M31. In this orientation, north is up, east to the left.
  The 25 Project AMIGA sight lines have red outlines; the supplemental
  sight lines have black outlines and are smaller. Two probable MS
  detections are filled in cyan. The remainder of the sight lines have
  non-detections of \HI\ associated with M31. Dotted circles show
  impact parameters $\rho = 100$, 200, 300 kpc, the last being roughly
  \Rvir. The sizes and orientations of the two galaxies are taken from
  the RC3 \citep{de-vaucouleurs1991} and correspond to the optical
  $R_{25}$ values. The dashed line shows the plane of the Magellanic
  System ($b_{\rm MS} = 0^\circ$) as defined by
  \cite{nidever2008}. The region within $b_{\rm MS} = \pm15^\circ$ is
  shaded. \label{fig:map}}
\end{figure}

We used the 100-m diameter Robert C. Byrd Green Bank Telescope
\citep{prestage2009} to take pointed observations toward \qsonumber\
AGNs projected behind the CGM of M31 as part of NRAO programs
GBT15A-328 and GBT14B-436. A map showing the locations of the AGNs is
given in Figure \ref{fig:map}, and the details of the AGN sight lines
are summarized in Table \ref{tab:targets}. These sight lines represent
two samples that, together, should provide an unbiased sampling of the
\HI\ content of the Andromeda galaxy's gaseous halo. The first sample
consists of 25 objects in the Project AMIGA sample, shown with red
outlines in Figure \ref{fig:map}. These are selected to be UV-bright
AGNs projected within $\rho \approx 340$ kpc ($\sim 1.1 \Rvir$) of the
center of M31. They are chosen to probe a broad range of impact
parameters, with sight lines focused about the major axis, minor axis,
and intermediate orientations. They do not sample the impact parameter
space randomly (i.e., with weighting $\propto \rho^2$). Furthermore,
they do not randomly sample the azimuthal distribution, both because
of the goals of Project AMIGA (to probe the azimuthal variations
systematically) and of a general dearth of identified UV-bright AGNs
behind the northern half of M31's CGM \citepalias[see][]{lehner2015}.

%%%%%%%%%%%%%%%%%%%%%%%%%%%%%%%%%%%%%%%%%%%%%%%%%%%%%%%%%%%%%%%%%%%%%%
%%%%%%%%%%%%%%%%%%%%%%%%%%%%%%%%%%%%%%%%%%%%%%%%%%%%%%%%%%%%%%%%%%%%%%

The remaining 23 pointings (shown with black outlines in Figure
\ref{fig:map}) sample sight lines toward UV-bright AGNs (and one blank
sky direction) that are not scheduled for \hst\ observations. These
supplemental sight lines were chosen because they may eventually
represent quality target for UV follow-up, and because they provide an
expansion of our survey in a way that mimics the initial selection. In
both cases, the principal criterion for selecting a sight line is the
presence of a UV-bright background source. This selection is
independent of the \HI\ distribution, and we used no knowledge of the
\HI\ structure of M31 in selecting background sources. There is a bias
in the distribution of the sources to the southern half of Andromeda's
CGM. This is in part due to the increased extinction northward of M31
(at lower Galactic latitudes) and a lack of surveys at those lower
latitudes \citep[a result being rectified by QSO searches with, e.g.,
LAMOST; ][]{huo2013}. We note that if the goal of our program was simply
to derive information about the covering factor of low column density
\HI\ in the vicinity of M31, we would not need to search in directions
toward background AGNs. However, the present program was designed with
the UV observations (or potential future observations) in mind, so
that deep \HI\ observations might help constrain the metallicity of
the CGM viewed in UV absorption.

The pointed GBT observations were taken with the L-band receiver and
have a \beamsize\ primary beam. Each AGN direction was initially
observed for a total of 50--60 minutes in individual 10 minute
scans. In a few cases difficulties with the receivers caused one of
the polarizations to be unusable. We received additional director's
discretionary time (program GBT16A-433) to reobserve 12 of the sight
lines.  Our observations employed a 5.16 MHz bandwidth centered on the
\HI\ hyperfine transition and have a native channel spacing of 0.15
\kms . The typical system temperature of the GBT in this configuration
is 18 K at zenith in both linear polarizations. Our observations are
consistent with this expectation, with most values in the range
$17.3 \la T_{\rm sys} \la 19.0$ K during our observations. The
exceptions are those observations taken toward radio-bright AGNs 3C48
(45 K) and 3C59 (21.9 K) as well as the direction toward 3C66A (24.6
K). None of the other background AGNs is within a factor of 10 of the
brightness of 3C48 and 3C59 (which have 1.4 GHz fluxes of $\sim 16$
and 2 Jy). In-band frequency switching (with a throw of 3 MHz) was
used for background subtraction. This provided useable velocity
coverage of typically $-515 \la \vlsr \la +470$ \kms.

\begin{figure}
% %run pyplot_spectra.py
\epsscale{0.75}
\plotone{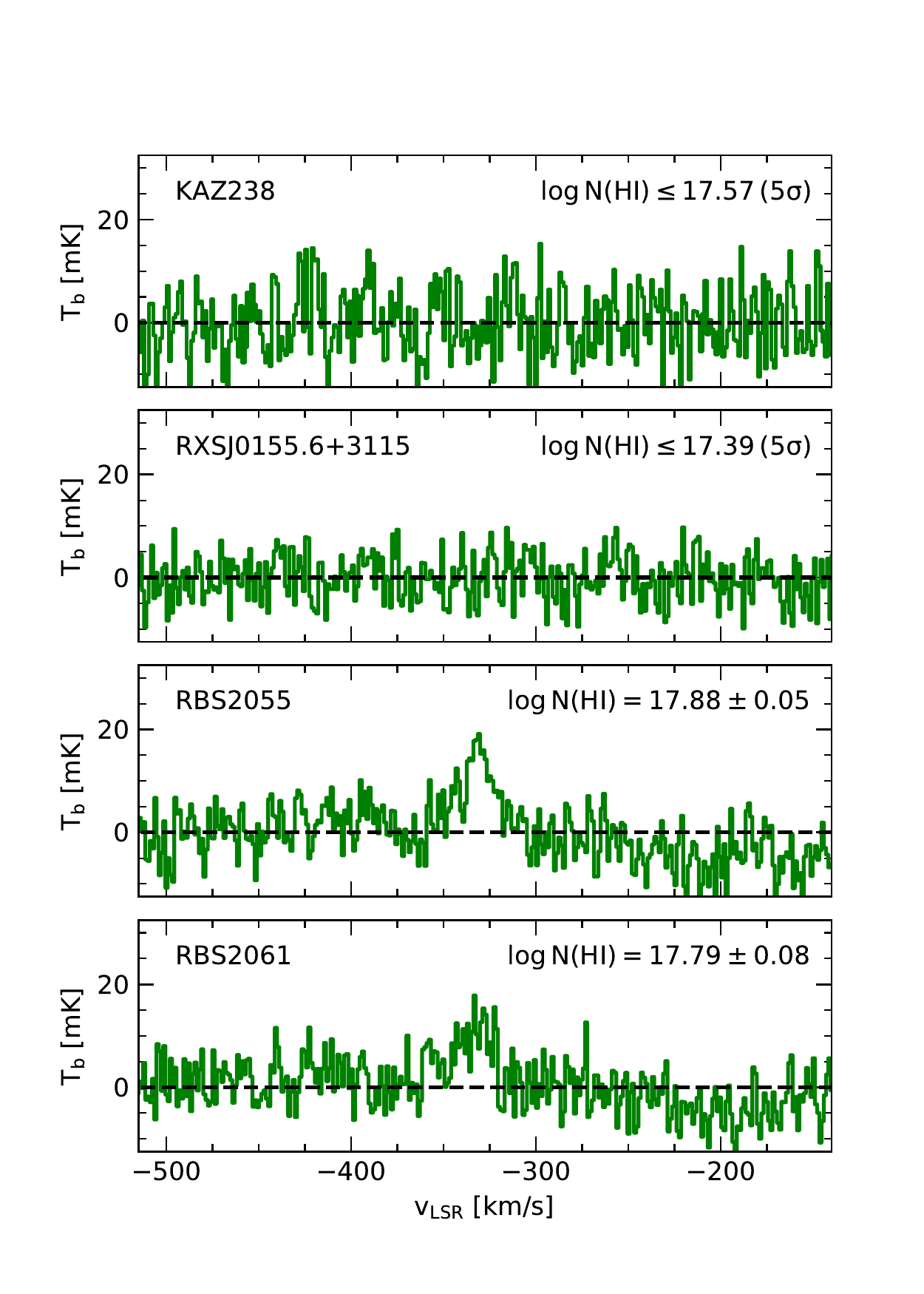}
\caption{Four spectra from our GBT observations covering the velocity
  range $-515 \le \vlsr \le -170$ \kms\ (sight line identifications
  are given in the upper left of each panel). The top two spectra are
  characteristic examples of the vast majority of the observations,
  showing no detectable \HI\ 21-cm emission from the CGM of M31. The
  bottom two spectra show the \HI\ emission observed toward RBS~2055
  and RBS~2061 (properties of the lines are given in Table
  \ref{tab:MSdetections}). All spectra are shown Hanning-smooted to
  $\Delta v\approx1.2$ \kms. Limits are determined using the RMS
  brightness temperature variations across the uncontaminated spectral
  range of the observations and assuming FWHM$\, = 25$
  \kms.  \label{fig:spectra}}
\end{figure}

We performed the data reduction within GBTIDL \citep{marganian2006},
following \citet{boothroyd2011} for the basic brightness temperature
calibration and stray radiation correction and \citet{lockman2012} for
the scan coaddition and baseline fitting. For each sight line we
individually examined each 10 minute scan, including both
polarizations. In some cases we interpolated over occasional localized
interference, i.e., for interference affecting a small number of
channels in a given scan. In a few cases, individual scans were
excluded from the coaddition if such interference occupied a large
number of channels. The quality of the spectral baselines is one of
the limiting factors in setting our column density sensitivity. For
some scans, one of the linear polarized receivers had much worse
baselines than typical.  We excluded these from the subsequent data
processing. We fitted spectral baselines separately to each scan and
linear polarization over a very broad velocity range before
coaddition. We adopted third- to fifth-order polynomials, taking care
to exclude regions of potential emission from the fitting. The
individual baseline-corrected spectra for a given sight line were then
coadded with equal weights after applying an atmospheric extinction
correction to each. Finally we corrected for the GBT's main beam
efficiency at 21-cm ($\eta_{\rm mb} = 0.88$). The final data are
binned to $\sim0.6$ \kms\ channel width. Several examples of our final
spectra are shown in Figure \ref{fig:spectra}.

We have performed an automated search for emission at $\ge 5\sigma$
significance over the velocity range $-515 \le \vlsr \le -170$ \kms\
(M31 has a systemic velocity $v_{\rm sys} = -300$ \kms ). Our
observations have typical RMS brightness temperature fluctuations of
$\sigma_{\rm b} \approx 8$ mK over the search velocities, with a full
range between $\approx 7$ and 12 mK per 0.6 \kms\ channel. The RMS
brightness temperatures for each sight line, derived empirically over
the full range of velocities searched for M31 emission, are given in
Table \ref{tab:targets}. Because we have calculated these empirically
they include both the random noise and the effects of imperfect
baseline subtractions or local baseline irregularities. 

For each sight line we calculate detection limits for the \HI\ column
density assuming the optically thin approximation:
$\NHI = [1.8 \times 10^{18} \ \mbox{\column\ (K \kms)}^{-1}] \ \int T_b
\, dv$. Any \HI\ in the CGM of M31 will be optically thin in the 21-cm
line.  The opacity of the 21-cm line is
$\tau = (5.2 \times 10^{-19} \ \mbox{K \kms\ cm$^2$}) \, \NHI \, {\rm
  FWHM}^{-1} \, T_{ex}^{-1}$ \citep{dickey1990}, where the excitation
temperature is generally close to the kinetic temperature of the \HI.
Thus for the expected \HI\ column densities $\approx 10^{18}$ \column,
FWHM$\, = 25$ \kms, and $T_{ex} > 1000$ K, the peak optical depth will
be $\tau < 2 \times 10^{-5}$.  Although the 21-cm measurements were
made toward AGN, these almost always contribute a negligible amount of
radio continuum emission.  Even for the exception, 3C48, which has a
continuum antenna temperature $\sim30$ K at the GBT, the continuum
temperature is much less than the 21-cm excitation temperature.  It
would reduce the 21-cm emission line by $<1$ mK for the expected 21-cm
opacities, a signal that is well below the noise level.

Our detection limits are given in Table \ref{tab:targets} based on the
RMS brightness temperature fluctuations and an assumed line width of
25 \kms\ (FWHM). This yields typical 5$\sigma$ detection limits of
$\NHI \le 4\times10^{17}$ \column\ per beam or better. Our choice of
FWHM is based on the median FWHM in detections of M31 HVC and clump
emission from the studies of \cite{thilker2004},
\citet{westmeier2008}, \citet{lockman2012}, and \citet{wolfe2016}.
Our typical sensitivity is equivalent to an \HI\ mass of
$M_{\rm HI} \sim 10^4$ \msun\ per beam ($\sim 800$
\msun$/{\rm kpc}^2$), no matter the scale of the emission.

%%%%%%%%%%%%%%%%%%%%%%%%%%%%%%%%%%%%%%%%%%%%%%%%%%%%%%%%%%%%%%%%%%%%%%
%%%%%%%%%%%%%%%%%%%%%%%%%%%%%%%%%%%%%%%%%%%%%%%%%%%%%%%%%%%%%%%%%%%%%%

% Although emission can be seen in two of the spectra (see Figure
% \ref{fig:spectra}), in no case do we definitively detect \HI\ 21-cm
% emission from the CGM of Andromeda. The two directions in which \HI\
% emission is observed at $\vlsr \approx -336$ \kms\ are associated with
% the extension of the Magellanic Stream (MS) in this direction.

Toward the direction of M31's extended halo, \HI\ emission from gas
associated with the Milky Way and its HVCs, the Magellanic Stream
(MS), and M31 can overlap in velocity, as discussed in detail by
several works \citepalias[\citealt{thilker2004},
][\citealt{kerp2016}]{lehner2015}.  M31 itself has a systemic velocity
in the LSR frame $\vlsr({\rm M31}) = -300\pm4$ \kms, and we expect
potential CGM emission within $\sim \pm225$ \kms\ of the systemic
velocity, given that more than 90\% of the candidate dwarf galaxies in
the M31 system lie within this range \citep{mcconnachie2012}.  The
upper velocity limit for our search is set by the likelihood of
contamination by emission from relatively local Galactic HVCs, which
are confined to $\vlsr \ga -170$ \kms\
\citep[][\citetalias{lehner2015}]{lehner2011a,lehner2012}. The lower
velocity limit corresponds to the limits of the M31 HVC emission found
by \citet{thilker2004}. None of the emission detected in high-fidelity
GBT individual pointings or maps \citep{thilker2004, lockman2012,
  wolfe2013, wolfe2016} have found emission at velocities more
negative than $\vlsr = -515$ \kms; neither have absorption line
searches for gas associated with the Andromeda galaxy
\citepalias[\citealt{rao2013}]{lehner2015}. In addition, most of our
spectra do not give access to velocities more negative than
$\vlsr \la -515$ \kms\ given the adopted frequency-switching
offset. At the same time, we can demonstrate that no Ly$\alpha$
absorption exists in the range $-1000 \le \vlsr \le -600$ \kms\ toward
the 25 AGNs in the Project AMIGA sample. This is based on a careful
search of the blue edge of the Milky Way Ly$\alpha$ damping
wings. Given the strength of this line, we would detect any gas with
$\logNHI \ga 13.5$. This search will be presented in a future paper.

In addition to potential contamination by Milky Way HVCs, the MS
crosses through this region of the sky at velocities that contaminate
searches for M31 emission, as detailed by \citetalias{lehner2015}
\citep[see also ][]{nidever2008, fox2014}. The emission from the MS is
most important at small Magellanic latitudes ($b_{\rm MS}$), although
metal ion absorption from the MS can be seen to at least
$b_{\rm MS}\approx\pm30^\circ$ \citep{fox2014} and at Magellanic
longitudes $l_{\rm MS} \ga -110\degr$. We show in Figure \ref{fig:map}
the position of $b_{\rm MS} = 0\degr$ as a dashed curve, and we shade
the region between $b_{\rm MS} = \pm15\degr$. The MS velocities can be
predicted following \citet{nidever2008}, as described in detail by
\citetalias{lehner2015} (see their Figure 3). Such contamination is
only important at $\vlsr \ga v_{\rm sys}({\rm M31}) \approx -300$
\kms\ \citep{de-vaucouleurs1991}.

Significant \HI\ emission over $-515 \le \vlsr \le -170$ \kms\ is
observed in only 2 of our 48 sightlines (see Figure \ref{fig:spectra})
at $5\sigma$ significance. The sight lines toward RBS~2055 and
RBS~2061 (the filled points in Figure \ref{fig:map}) show \HI\
emission at the level of $\logNHI \sim 17.8$. The \HI\ profiles for
these directions are shown in Figure \ref{fig:spectra}, and the line
properties are summarized in Table \ref{tab:MSdetections}. These
objects lie only $1\fdg2$ apart on the sky; both directions show
emission at $\vlsr \approx -336$ \kms. RBS~2055 and RBS~2061 have
Magellanic coordinates $(l_{\rm MS}, b_{\rm MS}) = (-113.2, +11.0)$
and $(-112.1, +11.6)$, respectively. Thus, both sight lines lie close
to the great circle on which the Magellanic system resides, and the
detected emission has velocities close to those expected for the
extrapolation of the MS in these directions. Our Project AMIGA COS
observations of the sight line towards RBS~2055 also reveal
significant metal line absorption. In particular, and unique among the
Project AMIGA sight lines, this sight line shows absorption from \OI\
$\lambda1302$ at the same velocity as the \HI\ emission (as well as
other metal lines). As we show in the Appendix, the comparison of
$N(\mbox{\OI}) / \NHI$, which gives a good measure of O/H with very
small if any ionization corrections, yields a metallicity estimate
[O/H]$\, = \MSmetallicity$. This is consistent with the metallicities
of the main body of the MS derived by \citet{fox2010, fox2013}.  This
agreement in metallicity with the general MS, and the location of
these directions relative to the MS, strongly suggests the detected
21-cm emission arises in the extension of the MS across this
region. We will assume these directions probe MS gas at
$\vlsr \approx -336\pm15$ \kms\ moving forward. We note that no \HI\
emission is detected in these directions outside of those expected for
the MS, although we do find absorption from M31's CGM at velocities
distinct from the MS in the strongest lines covered by the Project
AMIGA COS data (see the Appendix).

\section{\HI\ Column Density and Covering Factor Limits}
\label{sec:coveringfactor}

\begin{figure}[b]
% %run pyplot_nhirho.py
\plotone{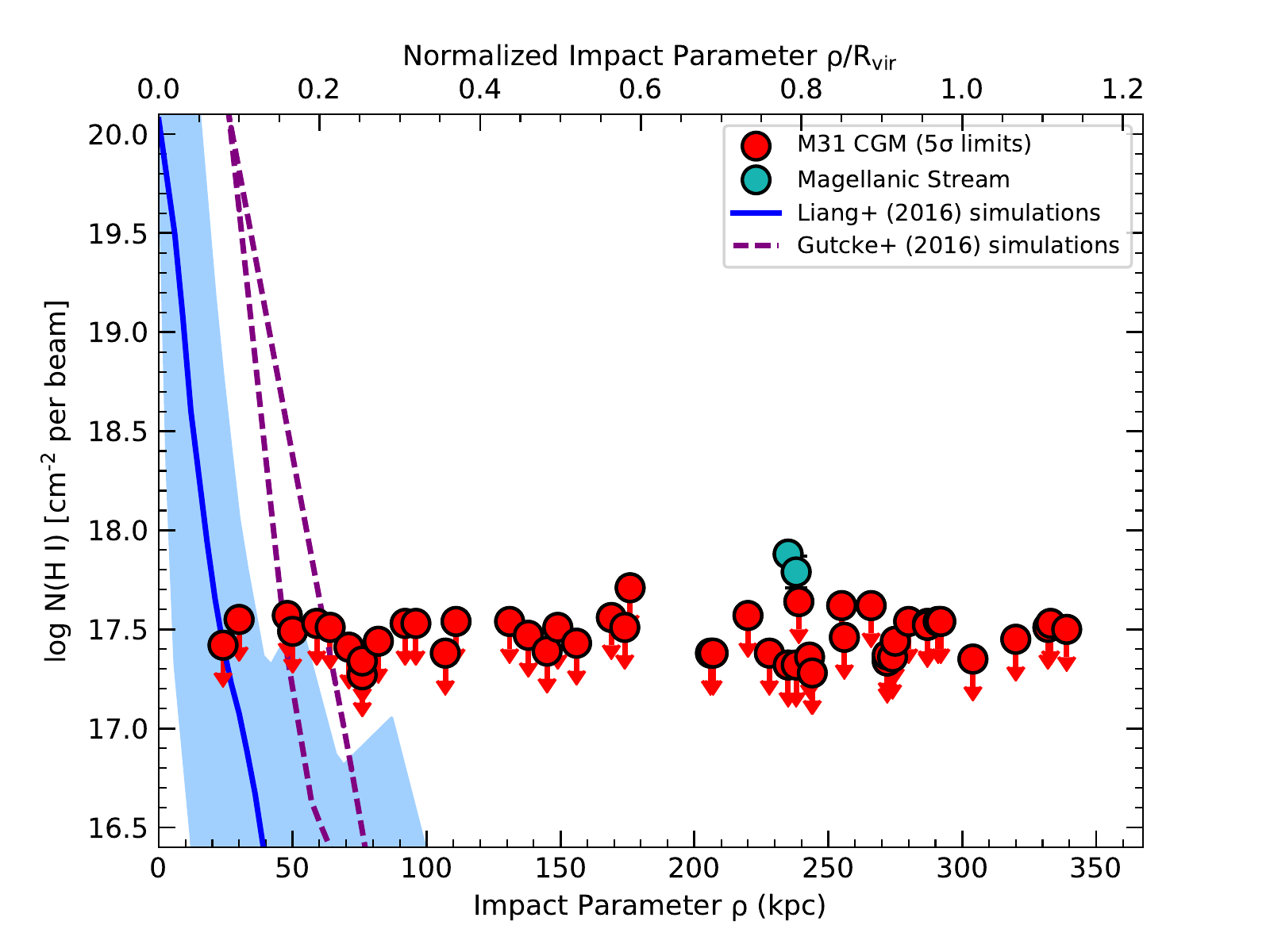}
\caption{Observed column densities and limits along each of the
  \qsonumber\ GBT sight lines through the CGM of M31. Upper limits
  (red) are given at the $5\sigma$ level assuming a ${\rm FWHM} = 25$
  \kms. They are determined over a velocity range
  $-515 \la \vlsr \la -170$ \kms. Two detections of apparent MS
  emission are shown in cyan. Those two sight lines would have
  $\log N({\rm H I}) \la 17.6$ outside of the velocities expected for
  the MS ($-370 \la \vlsr \la -300$ \kms). Also shown are profiles
  from recent simulations of \HI\ around massive galaxies. The blue
  line shows the median \HI\ profile from the $z=0$ simulation of
  \citet{liang2016} that best fits the ensemble of low-redshift CGM
  observations (their {\tt ALL\_Efb\_e001\_5ESN} run) along with the
  95\% confidence range of their values (cyan shading). This
  simulation follows a single $\log M_{\rm h}/M_\odot \sim 12$ halo to
  $z=0$; while it does fairly well at matching CGM observations, its
  stellar component is ``unrealistic'' \citep{liang2016}. The dashed
  lines are examples of results from \citet{gutcke2016}, showing the
  median values for their $\log M_*/M_\odot = 10.66$ and 10.89
  galaxies.  \label{fig:columnrho}}
\end{figure}

We have in hand a sample of \qsonumber\ pointings through the halo of
the Andromeda galaxy at impact parameters $\rho < 1.2 \Rvir$ with
sensitivity to \HI\ 21-cm emission of
$\NHI \approx 2 - 4\times10^{17}$ \column\ per beam. The \NHI\ limits
for each sight line are shown as a function of impact parameter in
Figure \ref{fig:columnrho} and reported in Table \ref{tab:targets}. To
encompass the range of sensitivites, we adopt a survey search
sensitivity of $\NHI \le 4\times10^{17}$ \column\
($\logNHI \le 17.60$).  The $5\sigma$ upper limits are shown for each
sight line, with the exception of the two MS detections discussed in
\S \ref{sec:observations}, for which the detected MS emission column
densities are shown. Outside of the velocity range
$-370 \la \vlsr \la -300$ \kms, these two sight lines show no emission
to limits similar to those observed along the other sight lines (see
Table \ref{tab:targets}).

%%%%%%%%%%%%%%%%%%%%%%%%%%%%%%%%%%%%%%%%%%%%%%%%%%%%%%%%%%%%%%%%%%%%%%

%%%%%%%%%%%%%%%%%%%%%%%%%%%%%%%%%%%%%%%%%%%%%%%%%%%%%%%%%%%%%%%%%%%%%%

Clouds with \HI\ column densities $\logNHI \ge 17.6$ could be present
on scales smaller than the 2 kpc diameter GBT beam and remain
undetected, as our limits apply to the beam-averaged column
density. Indeed, high-resolution observations of HVCs in the inner
regions of the halo ($\rho \la 50$ kpc) by \citet{westmeier2005} have
demonstrated that at least some M31 HVCs have smaller sizes
($\langle D \rangle \sim 1$ kpc) and higher central column densities
($\langle \NHI \rangle \approx 4\times10^{19}$ \column). Objects like
these specific clouds would be readily detected by our observations on
the basis of their masses ($\langle M_{\rm HI} \rangle \approx 10^5$
\msun), but there may still be small-scale clouds that lie below the
detection limits of our data. In fact, the photoionization simulations
by \cite{lehner2013} suggested relatively small sizes for the $z\la1$
LLSs in that work (the vast majority of which have
$N(\mbox{\HI}) \la 10^{17}$ \column), with only 11/25 of the systems
they modeled having sizes $>1$ kpc. We will discuss this issue further
below.

Also shown on Figure \ref{fig:columnrho} are predictions from the
recent ``zoom-in'' galaxy simulations. The blue curve shows the median
\HI\ profile from the {\tt ALL\_Efb\_e001\_5ESN} of \citet{liang2016},
which they note best fits recent COS observations of the CGM about
$z\la1$ galaxies while producing ``unrealistic'' results for the
stellar component of a typical low-redshift galaxy. The cyan shading
about the line represents the 95\% confidence range of the
measurements drawn from their simulation. Similar results from the
NIHAO simulations of \citet{gutcke2016} are shown in the dashed
lines. These represent the median profiles of their mock \HI\
distributions for individual galaxies with stellar masses
$\log M_*/\msun = 10.66$ and 10.89, those closest to M31. (Here we do
not show the quantiles, but they are similar in magnitude to those
shown for the \citeauthor{liang2016} simulations.)  The main point in
this comparison of both the \citeauthor{liang2016} and
\citeauthor{gutcke2016} simulations with our results is that \HI\
about a typical simulated galaxy falls off quite rapidly, save for the
filaments and smaller-scale structures tracing flows through the CGM.

The principal result of our work is a calculation of the cumulative
covering factor of \HI\ as a function of impact parameter from M31.
We calculate the covering factor from our data assuming a binomial
distribution. We follow \citet{cameron2011} in assessing the
likelihood function for values of the covering factor, \fccum, given
the number of detections (successes) against the total sample (number
of sight lines within a given impact
parameter). \citeauthor{cameron2011} demonstrates that the normalized
likelihood function useful for calculating (Bayesian) confidence
intervals on a binomial distribution with a non-informative (uniform)
prior follows a beta distribution.  We refer the reader to
\citet{cameron2011} for a detailed justification of this approach,
which is particularly useful for providing robust results in cases of
small samples (particularly important at small impact parameter in our
work).

%%%%%%%%%%%%%%%%%%%%%%%%%%%%%%%%%%%%%%%%%%%%%%%%%%%%%%%%%%%%%%%%%%%%%%
\begin{figure}
% %run pyplot_coveringfactor.py
  \plotone{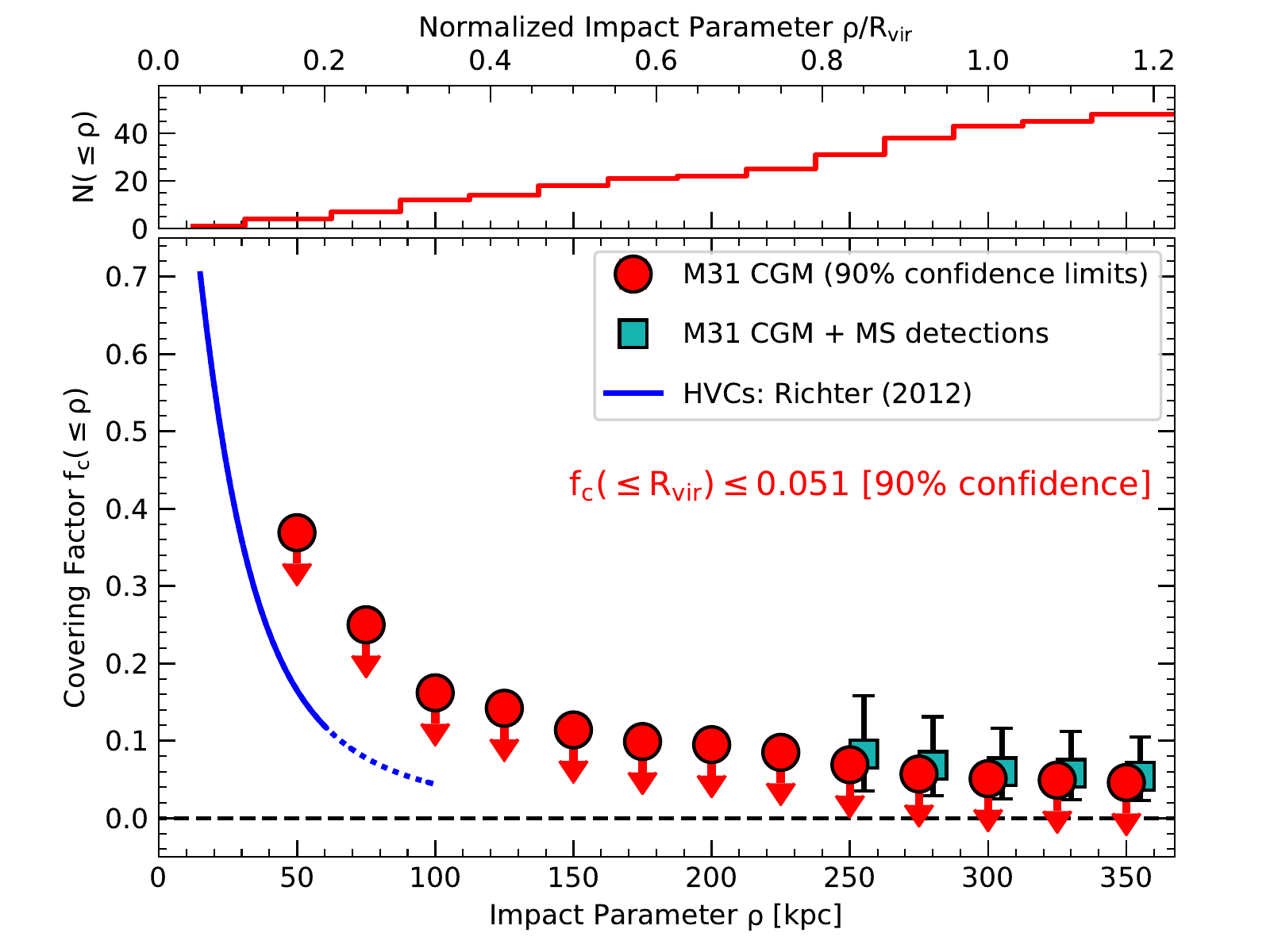}
  \caption{The cumulative covering factors for impact parameters less
    than $\rho$, \fccum, from our GBT observations. These values are
    appropriate for $\log N({\rm H I}) \ga 17.6$
    ($N(\mbox{\HI}) \ga 4\times10^{17}$ \column). Upper limits are
    shown at 90\% confidence. We assume a binomial distribution for
    each bin and follow the approach of \citet{cameron2011} in
    calculating the one-sided confidence limit using the incomplete
    beta function estimator (for a number of targets given in the top
    panel). The covering factor limit for impact parameters less than
    the virial radius is $f_c(\le \Rvir) \le \virialCovering$. Also
    shown in cyan are the covering factors that would be derived if
    one assumes the two sight lines with apparent MS emission are
    instead tracing emission from the CGM of M31 (shifted to slightly
    larger impact parameter for clarity). The central value is the
    median of the posterior probability distribution from the beta
    distribution generator, while the error bars denote the 80\%\
    confidence interval (i.e., the 10\%\ and 90\%\ quantiles)
    \citep{cameron2011}. The blue line shows the covering factor of
    HVCs in the inner regions of the halo (limiting sensitivity of
    $\logNHI \sim 18.25$), adjusting the results of
    \citet{richter2012} to reflect a cumulative covering
    factor. \label{fig:coveringfactor}}
\end{figure}
%%%%%%%%%%%%%%%%%%%%%%%%%%%%%%%%%%%%%%%%%%%%%%%%%%%%%%%%%%%%%%%%%%%%%%

The results of our covering factor calculations are given in Table
\ref{tab:coveringfactor} and shown in Figure
\ref{fig:coveringfactor}. We report {\em cumulative} covering factors:
each point represents the covering factor for all impact parameters
less than the one considered, \fccum, rather than the differential
covering factors (i.e., we calculate the covering factors for
$\le \rho$ rather than for $\rho \pm \Delta \rho$). Given the lack of
detections, we report one-sided confidence limits (the 90\% quantile
of the distribution) in Table
\ref{tab:coveringfactor}.\footnote{Specifically, we've used
  \citeauthor{cameron2011}'s Equation 3, determining the value of
  \fccum\ (equivalent to that equations's upper bound $p_u$) for which
  the integration of the beta distribution between \fccum\ and 1
  yields a normalized probability of 0.1.} We emphasize that the
results for each impact parameter are not independent, as they include
the same measurements as those at smaller impact parameter. Thus the
results summarized in Figure \ref{fig:coveringfactor} do not give
information on the {\em distribution} of covering factors within M31's
halo, only on the 90\% confidence limits to the covering factor within
a given impact parameter. [For interested parties, the limits on the
{\em differential} covering factors are $f_c^\prime \le 0.162$, 0.188,
0.099 at 90\%\ confidence for 100 kpc-wide bins centered on
$\rho = 50$, 150, 250 kpc, respectively.] We also note that our
covering factors are only appropriate for regions outside of the \HI\
disk of M31 (i.e., in the CGM). We define positions associated with
the disk as those that would fall on a projected circle of radius
$r=30$ kpc at an inclination angle of $i\approx 78^\circ$
\citep{braun1991}.

We also show in Figure \ref{fig:coveringfactor} the results we would
obtain if we assume the two \HI\ detections are in fact associated
with the CGM of M31, although we stress that this seems highly
unlikely (see Appendix). The covering factors derived under this
assumption are summarized in Table \ref{tab:MScoveringfactor} at
impact parameters for which the results are different than those in
Table \ref{tab:coveringfactor}. We give three values in Table
\ref{tab:MScoveringfactor}: \fccum $_{0.1}$, \fccum $_{0.5}$, \fccum
$_{0.9}$, which correspond to the 10\%, 50\%\ (median), and 90\%\
quantiles for the covering factor distributions \citep[again using the
incomplete beta generator discussed by ][]{cameron2011}.

We find the covering factor within \Rvir\ is
$f_c(\le \Rvir) \le \virialCovering$ (90\%\ confidence limit). Our
ability to provide meaningful constraints is dependent on the number
of sight lines probed, and thus the limits are not strong in the inner
regions where $<10$ sight lines were observed (see Figure
\ref{fig:coveringfactor}, top).  Even though we do not detect \HI\
emission in our survey, one should not be under the impression that
there is {\em no} high column density gas associated with M31's CGM,
as \citet{wolfe2016, kerp2016, westmeier2005, westmeier2008} have
found clumps of high column density gas in the halo (in some cases to
$\rho\sim100$ kpc from the center of M31). However, any such gas must
have a very small covering factor at, e.g., $\rho \ga 50$
kpc. \citet{westmeier2008}, for example, find $\sim95\%$ of the HVCs
in their map of a region of the M31 CGM are confined to $\rho \la 50$
kpc (for $M_{\rm HVC} \la 1.3\times10^5$ \msun\ at $5\sigma$). In our
data, gas at column densities $\logNHI \ga 17.6$ can be present on
scales smaller than the 2 kpc-diameter beam and remain undetected, so
long as the total mass is less than our typical sensitivity of
$M_{\rm HI} \approx 10^4$ \msun.

Figure \ref{fig:coveringfactor} also shows an assessment of the
covering factor of HVCs about M31 by \cite{richter2012} \citep[based
on the maps of][]{thilker2004}. The blue curve in Figure
\ref{fig:coveringfactor} shows these results rescaled for our assumed
distance and transformed into the cumulative covering factors used
here based on \citeauthor{richter2012}'s fit. The results beyond
$\rho \approx 60$ kpc are an extrapolation of that fit. This fit
applies only to the HVC component as defined by \cite{thilker2004}. It
excludes the disk of M31. Our new limits are compatible with the HVC
covering factors, although the HVCs were observed over an area in
which we have few measurements and limited constraints. The HVCs (and
``bridge'' material) observed about M31 have typical column densities
in excess of $\logNHI \ga 18$ and masses $M_{\rm HI} \ga 10^5$ \msun,
which would be readily detectable by our observations if such material
were common in our survey area.  While we discuss our results in the
general context of the CGM, our definition of CGM is inclusive of
``high velocity clouds.'' Our limits show there is not an abundance of
HVCs beyond the mapping limits of those earlier works. A caveat is
that our covering factor determinations are limited to
$-515 \le \vlsr \le -170$ \kms, the lower end set by
frequency-switching limits and the upper end set to avoid
contamination by Galactic HVCs.

\section{Comparison with Recent CGM Surveys and Simulations}
\label{sec:comparisons}

\subsection{Comparison with Recent Observations}
\label{sec:obscomparison}

The nearest comparison to our measurements from the QSO absorption
line literature are those of the \coshalos\ survey
\citep{tumlinson2011, tumlinson2013, prochaska2017}. The original
survey targeted AGNs projected within $\rho \sim 150$ kpc of 44
$\sim$\lstar galaxies at $z\sim0.2$, with detection of \HI\
Lyman-series absorption in nearly every case \citep[40/44;
see][]{tumlinson2013}. Such absorption line observations can be
sensitive to column densities as low as $\logNHI \approx 13$. Most of
the \HI\ detections in the original survey had \HI\ column densities
that were either orders of magnitude below our detection limits or
that were not able to be determined due to the strong saturation of
the available Lyman series lines. Follow-up measurements of the Lyman
break in some of these these galaxies have recently been completed,
providing stronger constraints on their \HI\ column densities
\citep{prochaska2017}. We compare our results with this most recent
update.

\begin{figure}
  % % run pyplot_nhirhoCH.py
  % % run pyplot_coveringfactorCH.py
  % \epsscale{1.5}                %
  % \plottwo{fig-NHIrhoCH.pdf}{fig-coveringFactorCH.pdf} 
  %%
  % % run pyplot_CHcomparison.py
  \epsscale{0.75}
  \plotone{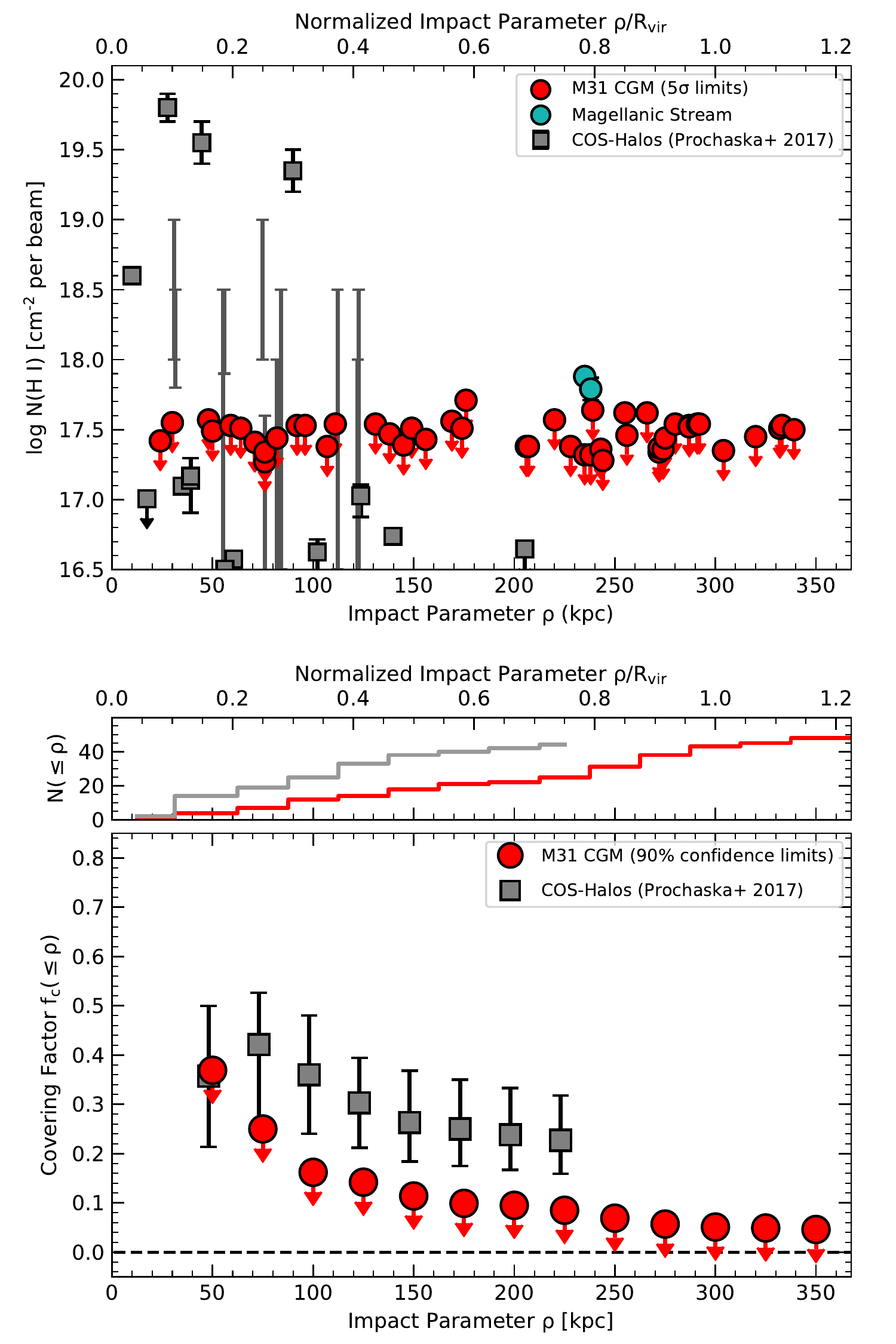}
  \caption{{\em Top:} The column density limits through the CGM of M31
    compared with the results of the absorption-line \coshalos\ survey
    \cite[see also \citealt{thom2012,
      tumlinson2013}]{prochaska2017}. The error bars without central
    values represent sight lines for which \cite{prochaska2017}
    bracket the column densities with upper and lower limit, arguing
    for a flat probability distribution between those 95\% confidence
    limits. Because the \coshalos\ results probe galaxies over 2 dex
    in stellar mass ($9.5 \la \log M_*/M_\odot \la 11.5$), we plot the
    position of each galaxy according to its normalized impact
    parameter, $\rho / \Rvir$ (i.e., according to the impact parameter
    scale on the top axis) for comparison with the M31 results. Only
    about half of the \coshalos\ systems are seen in this plot, the
    rest having column densities $\log \NHI \le 16.5$. All of the
    \coshalos\ galaxies are at $\rho / \Rvir \la 0.75$. {\em Bottom:}
    Covering factor limits for M31 compared with the covering factor
    of $\log \NHI \ge 17.6$ absorption in the ensemble of \coshalos\
    galaxies. Here we have calculated the cumulative covering factors
    for the revised \coshalos\ survey of \HI\ \citep[][see also
    \citealt{thom2012,tumlinson2013}]{prochaska2017}. The central
    value for the \coshalos\ results is the median of the output
    distribution from the Monte Carlo approach described in the text,
    while the error bars denote the 10\%\ to 90\%\ quantiles.  All of
    the \coshalos\ galaxies are at $\rho / \Rvir \la 0.75$, so we only
    plot covering factors within that limit. Note that the
    distribution of sight lines for the \coshalos\ sample (grey
    histogram in the upper panel) is significantly different than that
    of our Andromeda sample, with the former weighting smaller
    normalized impact parameters more
    heavily.\label{fig:CHcomparison}}
  %\label{fig:CHcomparison}}
\end{figure}

Figure \ref{fig:CHcomparison} compares the updated \coshalos\
measurements with our measurements in the halo of the Andromeda
galaxy. The top panel shows \NHI\ versus $\rho$ from
\citet{prochaska2017}. All of the \coshalos\ sight lines are
considered, though many of the \HI\ columns (19/44) fall below the
lower bounds of this plot. Several of the \coshalos\ sight lines have
only broad constraints, with lower limits from saturated Lyman series
lines or a saturated Lyman break and upper limits from the lack of
damping wings on \lya\ or \lyb. \citeauthor{prochaska2017} argue these
systems should be treated as having a flat probability distribution in
\logNHI\ between their lower and upper bounds. We adopt this
recommendation, plotting these systems in Figure
\ref{fig:CHcomparison} as extended error brackets (representing the
95\% confidence interval) without central values.

%%%%%%%%%%%%%%%%%%%%%%%%%%%%%%%%%%%%%%%%%%%%%%%%%%%%%%%%%%%%%%%%%%%%%%
%%%%%%%%%%%%%%%%%%%%%%%%%%%%%%%%%%%%%%%%%%%%%%%%%%%%%%%%%%%%%%%%%%%%%%

There are several caveats to note when comparing our M31 measurements
to those of \coshalos. The \coshalos\ sample probes a broad range of
galaxy mass/luminosity centered roughly on ``\lstar.''  The estimated
halo masses for the \coshalos\ survey cover the range
$11.5 \la \log M_{\rm h} / \msun \la 13.7$, with stellar masses
$9.6 \la \log M_* / \msun \la 11.5$; these should be compared with the
\citet{tamm2012} estimates for M31 of
$\log M_{\rm h} / \msun \approx 12.0$ and
$\log M_* / \msun \approx 11.0$. Thus, we are comparing one galaxy to
an ensemble of galaxies with a broad range in masses. Furthermore, due
to this broad mass range, the relevant physical scales can vary
significantly. We have chosen to present all \coshalos\ measurements
and results relative to their normalized impact parameter,
$\rho / \Rvir$, which is shown across the top of the
figures.\footnote{The results of our comparison are not fundamentally
  different if we consider the results with $\rho$ rather than the
  normalized coordinate.} For this reason some of the \coshalos\
measurements will appear at impact parameters larger than the survey
limit of $\rho \la 150$ kpc, as the more massive systems in \coshalos\
have virial radii larger than that of
M31. % Readers should also bear in mind that the \coshalos\
% absorption line measurements are made over a sub-parsec scale at the
% host galaxies. Thus, the effective ``beam'' of the absorption line
% results is quite different than our GBT beam, which subtends 2 kpc at
% Andromeda.

The bottom panel of Figure \ref{fig:CHcomparison} shows a comparison
of the covering factors derived for the \citet{prochaska2017} results
in comparison with ours for M31. The \coshalos\ results are based on a
Monte Carlo sampling of the full \NHI\ distribution over a given
(normalized) impact parameter range. We do this to account for the
systems with lower and upper bounds on the column density that
straddle our detection limit (though we have verified that it gives
results indistinguishable from the \citealt{cameron2011} approach if
we consider only systems clearly above our detection limit). For each
impact parameter considered, we create an \NHI\ distribution that is
the sum of the probability distribution functions (PDFs) describing
each of the \coshalos\ measurements. The direct measurements in
\citet{prochaska2017} have reasonably-symmetric errors, and we adopt a
normal distribution for each in \logNHI. For the bounded values from
\citeauthor{prochaska2017}, we follow their recommendation and treat
each as having a flat PDF between the limiting columns. For upper
limits, we assume a flat PDF between $\logNHI = 10.0$ and the quoted
upper limits. To calculate the covering factor within a given impact
parameter, we create 10,000 realizations of the observations, drawing
the same number of \HI\ column densities as sight lines within that
range from the full \NHI\ distribution. For each realization we
calculate the covering factor as the fraction of the mock sight lines
with $\logNHI \ge 17.6$. From the full sample of covering factors, we
derive the quantiles of the distribution. The bottom panel of Figure
\ref{fig:CHcomparison} shows the results, where the central values
give the median of the distribution with error bars representing the
[10\%, 90\%] quantiles (i.e., the 80\% confidence interval).

The implied covering factor of gas with $\logNHI \ge 17.6$ for
\coshalos\ galaxies within $0.5 \Rvir$ is $f_c(\le 0.5 \Rvir) = 0.26$
[0.18, 0.37] (median and [10\%, 90\%] quantiles), and it is
$\approx0.36$ within $0.33 \Rvir$. (Although they do not normalize to
the virial radius, these results are similar to those given by
\citet{prochaska2017}, who note the covering factor at $\rho \le 75$
kpc is $\ga 50\%$ for $\log \NHI \ga 17.2$.)  The \coshalos\ survey
thus shows a striking prevalence of high column density \HI\
absorption systems. A sizeable fraction of the \coshalos\ absorption
line measurements (at least 8/44) have columns in excess of our
detection limit, $\logNHI = 17.6$. The mean \NHI\ for gas with
$\logNHI \ge 17.6$ from \coshalos\ is
$\log \langle \NHI \rangle_{\ge17.6} = 19.15$ (with a geometric mean
$\langle \log \NHI \rangle_{\ge17.6} = 18.65$). This gas is detected
frequently enough that the mean \NHI\ drawn from our construction of
the full \NHI\ distribution of \citet{prochaska2017} is
$\log \langle \NHI \rangle = 18.51$.

In roughly 30\% of the \coshalos\ galaxies, gas with
$\log \langle \NHI \rangle_{\ge17.6} = 19.15$ is observed within
$0.5 \Rvir$. Any of our observations intercepting gas with this column
density would readily detect such gas if it covered 30\% of our 2 kpc
beams. Thus, if the mean \coshalos\ statistics are a fair
representation of the structure of \NHI\ on small scales --
specifically as good reprentations of the fraction of a 2 kpc diameter
beam covered by any high column density gas clouds -- we should have
detected any clouds having high \HI\ columns like those seen in the
\coshalos\ measurements. Indeed, gas with $\logNHI = 19.15$ would need
to cover $\le 2.8\%$ of our beam to avoid detection, some $10\times$
lower than implied by the \coshalos\ covering factors.

It is the case that the radial distributions of sight lines in
\coshalos\ and our sample are weighted differently. The high column
density \HI\ absorbers in \coshalos\ are at $\rho \la 0.33 \Rvir$
(Figure \ref{fig:CHcomparison}). Even considering only this range in
impact parameter, however, our M31 covering factors
$f_c(\le 0.33 \Rvir)$ aren't consistent with those in the \coshalos\
galaxies. Thus, it is difficult to reconcile the ensemble results from
\coshalos\ with our measurements of \HI\ in the CGM of the Andromeda
galaxy.

The \coshalos\ galaxies that show gas at $\log \NHI \ge 17.6$ have a
mean stellar mass $\langle \log M_*/\msun \rangle \approx 10.9$ (full
range 10.2 to 11.3), so this high column density gas resides in the
halos of galaxies similar to M31. It is perhaps noteworthy that the
\coshalos\ survey attempted to select against pairs of \lstar\
galaxies \citep{tumlinson2013}, such as the M31-Milky Way pair. Thus,
if M31 were at $z\sim0.2$, it may not have been been included in the
\coshalos\ survey. If its Local Group membership plays a role in
determining the covering factor, the M31/\coshalos\ comparison could
be inappropriate (although the halos in the Local Group-like
simulations of \citealt{nuza2014} show results similar to those of the
single-halo simulations discussed below).  However, because the
initial selection was done using photometric redshifts, this selection
against galaxy pairs was not as clear cut as initially intended
\citep{werk2012}.

% The results shown in Figure \ref{fig:CHcomparison} beg the question:
% are the covering factors of optically-thick gas about the ensemble of
% \coshalos\ galaxies consistent with those of our survey of the CGM of
% Andromeda? In short, no. Taking the ensemble of absorption line
% measurements from \citet{prochaska2017} as a model for the structure
% within the CGM of Andromeda, we would readily have detected 21-cm
% emission if the \coshalos\ results were representative of M31 (with
% $\langle \logNHI \rangle = 18.8$ and $f_c \approx 0.3$). This is true
% even if the CGM contains substructures that do not fill our beam. 

In general metal absorption lines show higher covering factors about
galaxies than our \HI\ 21-cm measurements about M31. Metal lines are
found with covering factors $>60\% - 75\%$ within $\sim0.5 \Rvir$
\citep[e.g.,][]{chen2010, nielsen2013, stocke2013, werk2014,
  borthakur2016}. Indeed, \citetalias{lehner2015} find covering
factors near unity for \CII\ and \SiIII\ absorption within
$\rho \la 0.5 \Rvir$, albeit with a very small sample. Preliminary
results from Project AMIGA still support a high covering factor in
these ions (N. Lehner et al. in prep). The metal lines show higher
covering factors in part due to the greater sensitivity of absorption
line techniques to low column density gas (e.g., \lya\ absorption
lines are detectable to $\logNHI < 12.5$), but it is also a reflection
of the general ionization level of gas in the CGM. A majority of the
cool and warm gas ($10^{4-5} \le T \le T_{\rm vir}$) in galaxy halos
is significantly ionized \citep[][]{werk2014,keeney2017}, including
the LLS-like regime that we are probing \citep{lehner2013,
  fumagalli2016, lehner2016}.

The metal ion covering factors about M31 are not that different than
those found about \coshalos\ and other galaxies. But, depending on the
degree of saturation within the detected metals, we should not
necessarily expect to see the same covering factor in high \HI\ column
density gas. We have compared the \citetalias{lehner2015} measurements
along the three sight lines with $\rho \la 50$ kpc relative to
M31 to the \coshalos\ results. All of the \coshalos\ with high
\HI\ column densities ($\logNHI \ge 17.6$) also have significantly
higher \ion{Si}{2}, \ion{Si}{3}, and \ion{C}{2} columns than the three
inner-halo Andromeda sight lines. These \coshalos\ systems have
$>2-3\times$ higher \ion{Si}{3} columns than M31 (all are lower limits
due to saturation in the \coshalos\ sight lines) and $\sim10\times$
higher \ion{Si}{2} column densities. Thus, it appears the \coshalos\
sight lines simply probe higher column densities of low-ionization
gas, including \HI. This is hidden by the comparison of metal ion
covering factors because the metal absorption lines are so sensitive
and quick to saturate. The metal / \HI\ ratios in M31 appear to be
consistent with those found by \coshalos.

\subsection{Comparison with Recent Simulations}

The covering factors of \HI\ and metal ions are considered in a number
of simulation papers \citep{fumagalli2011, cafg2011, shen2012,
  fernandez2012, fumagalli2014, cafg2015, suresh2015, cafg2016,
  gutcke2016, liang2016}. The motivation for extracting this quantity
from simulations arises in part because this is an observable quantity
\citep{rudie2012,prochaska2013}, but also because the covering factor
can respond to changes in the accretion rate and feedback intensity in
a galaxy and the numerical approaches adopted.  For example, the
covering factors of LLS absorption can change by factors of $>2$
depending on the wind models adopted or presence of AGN activity
\citep[e.g.,][]{suresh2015}. Most commonly the predictions are made
for $z\sim2$, not only because the simulations are less expensive to run
to these redshifts, but also because there are more measurements of
the \HI\ covering factor at high redshift than at $z\sim0$, which
typically requires space-based observations. At the same time, the
choice of the limiting \HI\ column for reporting the covering factors
varies widely, depending on the focus of the simulations.

\begin{figure}
% %run pyplot_coveringfactorModels.py
\plotone{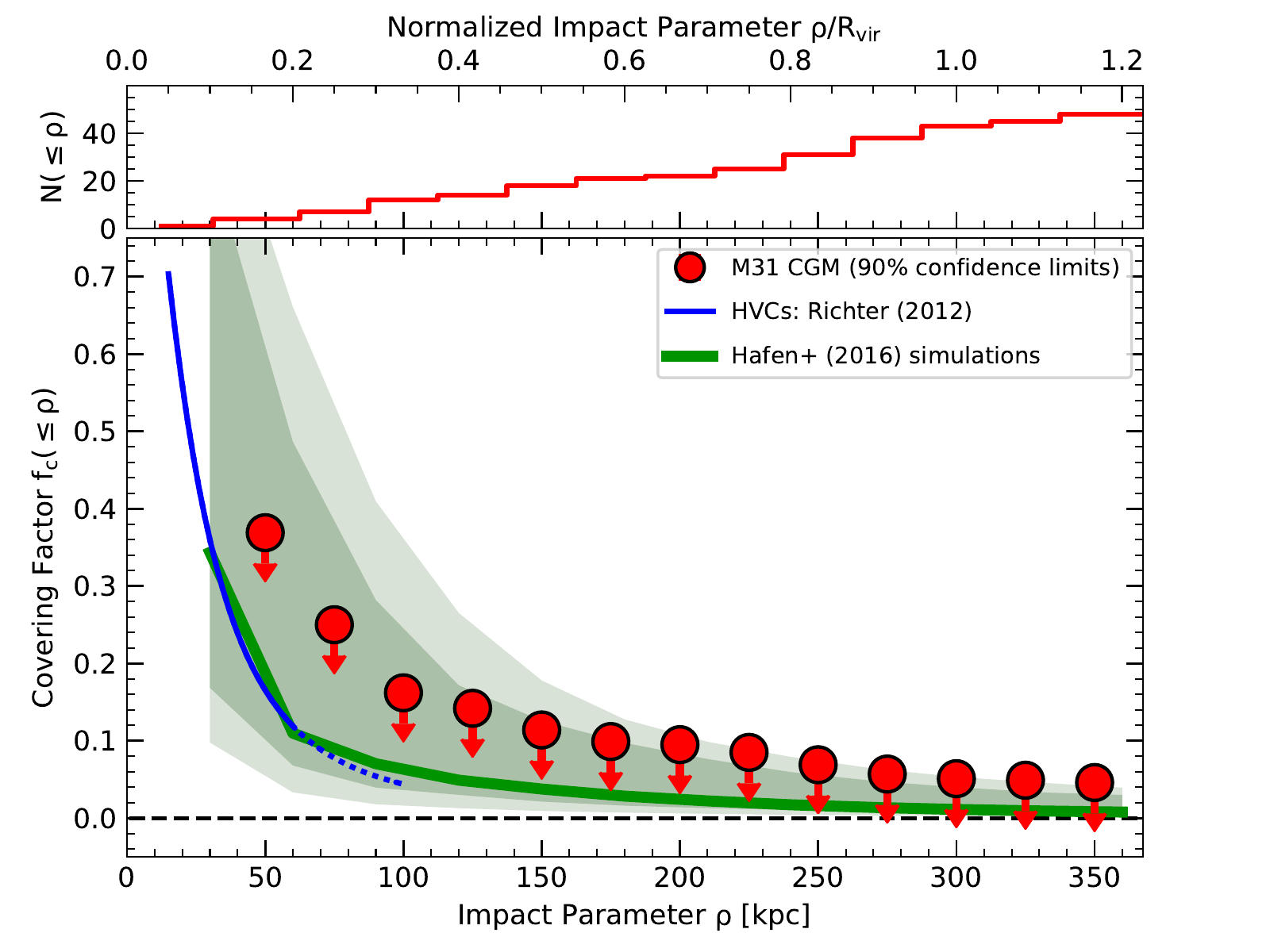}
\caption{The cumulative covering factors for M31 compared with
  simulation results.  The green line represents the median cumulative
  covering factor from the $11.8 \le \log M_{\rm h}/M_\odot \le 12.1$
  simulations of \citeauthor{hafen2017} \citepalias{hafen2017}, while
  the dark and light green shading shows the [25\%, 75\%]
  (interquartile) and [5\%, 95\%] ranges from these models. The median
  simulation result has a covering factor $f_c (\le \Rvir) = 0.011$.
  Our limits on the covering factor from M31 are consistent with
  expectations from these
  simulations.  \label{fig:coveringfactorModels}}
\end{figure}

With this in mind, we compare our observations of M31 with the recent
simulations reported in \citetalias{hafen2017}, who investigated the
relationship of $z<1$ LLSs to galaxies; these halos were simulated as
part of the FIRE project \citep{hopkins2014}. Figure
\ref{fig:coveringfactorModels} compares our results with an ensemble
of cosmological zoom simulations from \citetalias{hafen2017}, shown
with the green curve and shaded regions.  The analysis of
\citetalias{hafen2017} produce simulated column density distributions
for their ensemble of galaxies, which they found to be consistent with
the cosmological incidence rates of low-redshift ($0 \le z \le 1.0$)
LLSs at $\logNHI \ge 17.5$ and $\logNHI\ge 17.2$
\citep{ribaudo2011}. These simulations however have difficulties
matching the observed metallicity distribution of low-z LLSs
\citep{lehner2013, wotta2016}.  The results summarized in Figure
\ref{fig:coveringfactorModels} are from the four halos studied in
\citetalias{hafen2017} whose final ($z = 0$) masses are
$11.8 \le \log M_{\rm h}/M_\odot \le 12.1$ (simulations m11.9a, m12i,
m12q, m12v of \citetalias{hafen2017}). The simulated halos have been
projected into a $512^3$ grid spanning $2.4\times R_{\rm vir}$ (with
cell sizes $\sim1.0$--1.2 kpc) for each simulation.  We derived
cumulative covering factors for $\log N(\mbox{\HI}) \ge 17.6$ using
normalized impact parameters $\rho/R_{\rm vir}$ as we did in the
COS-Halos comparison. We consider each of the simulated galaxies from
three orthogonal perspectives and use 11 snapshots per simulation over
the redshift range $0 \le z \le 0.25$.  Thus there are 132 total
models that go into calculating \fccum\ and its quantiles for each
impact parameter bin (which are sampled in steps of
$0.1 \, \rho/\Rvir$).

%%%%%%%%%%%%%%%%%%%%%%%%%%%%%%%%%%%%%%%%%%%%%%%%%%%%%%%%%%%%%%%%%%%%%%
%%%%%%%%%%%%%%%%%%%%%%%%%%%%%%%%%%%%%%%%%%%%%%%%%%%%%%%%%%%%%%%%%%%%%%

The thick curve in Figure \ref{fig:coveringfactorModels} shows the
median of the \citetalias{hafen2017} results, the darkest shading
shows the interquartile range (from the 25\%\ to 75\%\ quantiles), and
the lightest shading shows the 5\% to 95\% quantile range.  The median
results show that the covering factor from these simulations of high
column density gas around galaxies in the mass range analyzed is
generally relatively small, with simulated median covering factors
$f_c (\le 0.5 \Rvir ) = 0.037$ and $f_c(\le R_{\rm vir}) = 0.011$.
These compare with our upper limits of $<0.11$ and $<0.051$ at 90\%
confidence for the same two impact parameters; they are significantly
lower than the results from \coshalos\ at these same impact
parameters. We point out that the simulations display significant
variations in the covering factors. The variations are a result of
both strong time-dependence related to wind and accretion activity (in
the case of the \citetalias{hafen2017}) as well as variations between
different galaxy halos \citep[][\citetalias{hafen2017}]{cafg2015,
  muratov2015}.

The highest covering factors in the simulations arise in the inner
regions of the galaxies. We do not sample $\rho \la 50$ kpc well with
our 21-cm observations due to our focus on directions with known
UV-bright AGN (we have only 4 directions in this regime). In the inner
regions (within $\rho \approx 0.33 \Rvir \approx 100$ kpc) the FIRE
simulations from \citetalias{hafen2017} seem to be in good
agreement with the covering factors derived for M31's HVC population
\citep{richter2012}. 

Other simulations find results similar to those of
\citetalias{hafen2017}. For example, \citet{gutcke2016} find
$f_c\approx 0\% - 10\%$ for $\logNHI \ge 17.2$ in simulations of
M31-like mass galaxies at $z\sim0$, where the range represents the
variations seen in simulations of different
galaxies. \citet{fernandez2012} presented a high-resolution simulation
of a single M31-like halo, and they found \HI\ covering factors
reasonably consistent with the \citet{richter2012} distribution for
$\logNHI \ge 17.7$. \citet{nuza2014} considered a Local Group-like
configuration of halos; they find very high covering factors in the
inner regions of those two halos, and results that are generally
consistent with our observations.  Our measurements in M31 are
consistent with the median set of simulations in all of these cases,
showing a quite low covering factor of optically-thick \HI\ when
assessed within \Rvir.

\section{Discussion}
\label{sec:discussion}

In this work, we have investigated the frequency of optically-thick
gas with $\log \NHI \ge 17.6$ about the local $\lstar$ galaxy M31. On
the basis of GBT searches for 21-cm emission with a 2~kpc beam, we
find a minimal covering factor of such gas in the halo of the
Andromeda galaxy. We derive a covering factor within the virial radius
of $f_c (\le \Rvir) < \virialCovering$. Our limits on the cumulative
covering factor are not particularly strong in the inner halo (esp.,
$\rho \la 0.25 \Rvir$), where we have few pointings. Complementary
information exists from the previous HVC results within
$\rho \la 0.2 \Rvir$ \citep{thilker2004, richter2012}.

There is some discrepancy in our assessment with previous work on the
\HI\ environment of the M31/M33 system. As noted earlier, the initial
maps by \citet{braun2004} seem to show more wide-spread emission than
is present \citep{lockman2012, wolfe2013, wolfe2016}.\footnote{These
  works generally show a very low detection rate in their searches,
  consistent with our low covering factor determinations. However,
  because these works studied areas specifically to address previous
  potential detections of \HI\ by \citet{braun2004}, their
  observations are not suitable for general covering factor
  calculations, as they may be biased in favor of \HI\
  detections. Nonetheless, these efforts all support a small covering
  factor on small scales, even in directions suspected originally to
  have \HI\ emission.} More recently, \citet{kerp2016} have emphasized
the potential for blending of high-velocity M31 structures with the
Milky Way's 21-cm emission. They attempted to decompose the
complicated region around M31 with guidance from continuities in the
``difference second moment map,'' arguing that several very large \HI\
complexes within $\rho \sim 80$ kpc are associated with M31. These
complexes span tens of kpc at the distance of M31. We would not have
included them as detections in our survey because they lie outside of
our velocity range. The \citeauthor{kerp2016} structures are at
$-150 \la \vlsr \la -100$ \kms\ (within $\approx200$ \kms\ of M31's
systemic velocity), whereas our survey is restricted to
$-515 \le \vlsr \le -170$ \kms\ in order to avoid contamination by
Galactic HVC emission. Furthermore, none of our pointings directly
intersect the clouds they identify, a by-product of our low sampling
density at small impact parameters. Assuming all of the clouds
\citeauthor{kerp2016} identify are indeed associated with M31, their
Figure 7 implies covering factors just consistent with our limits. For
example, their map implies $f_c(\le 0.33 \Rvir) \approx 0.2$ (ignoring
emission within 25 kpc in order to exclude the disk), which compares
with our limit $f_c(\le 0.33 \Rvir) \le 0.16$ at 90\% confidence. This
covering factor estimate is dominated by the two large structures that
they identify on opposite sides of the galaxy (their clouds {\em a}
and {\em b}). It is a bit disconcerting that the two structures,
separated by many tens of kpc, both lie at the extreme positive
velocities expected for circumgalactic material about M31 (both with
$\vlsr \approx -115$ \kms). It is not clear why two structures
separated by more than the diameter of M31's disk should have the
exact same extreme velocity. The covering factors implied by the
\citet{kerp2016} are $\sim2\times$ higher than those of
\citet{richter2012} at the same impact parameters. This discrepancy is
wholly connected to the issues of blending with Milky Way HVC gas at
$\vlsr \ga -170$ \kms. While \citeauthor{kerp2016} attempt to solve it
through a unique approach, it is difficult to verify their results
with other means.

In general our limits to the \HI\ covering factor are quite consistent
with the low values expected from cosmological zoom simulations of
individual galaxies at $z \approx 0$
\citep[][\citetalias{hafen2017}]{gutcke2016}. With no detections, we
do not have a valid characterization of the shape of the $f_c$
distribution about M31, although the work by \cite{richter2012}
provides a fit to the HVC covering factor within $\sim60$ kpc. Some of
the simulated galaxies show higher $f_c$ than is observed in M31, as
captured in the simulations ranges shown in Figure
\ref{fig:coveringfactorModels}. This is a result of the halo-to-halo
variations as well as the significant temporal variations within a
single halo \citep[][\citetalias{hafen2017}]{rahmati2015,
  cafg2015}. Thus, while the covering fraction of optically-thick gas
about M31 is not in conflict with the simulations, it is a single
sample from a specific time. It may be drawn from a broad distribution
like those seen in the simulations.

% And, we should not However, cold gas on small scales is the most
% difficult regime to produce correctly in such simulations, so we
% should view the agreement with caution.

As discussed in \S \ref{sec:obscomparison}, our results appear
discrepant from the \coshalos\ galaxies \citep{tumlinson2013,
  prochaska2017}. There are very few other works that put strong
constraints on the covering factor of optically-thick gas about
low-redshift galaxies. There could be a multitude of physical reasons
for the difference between M31 and \coshalos : 1) M31 could have
experienced significant evolution in the last 2.5 Gyr, the time since
the typical redshift of the \coshalos\ galaxies; 2) ``green valley''
galaxies with star formation rates like M31, which are not well
sampled by \coshalos, may have distinct CGM properties; 3) the
presence of M31 in a group with another massive spiral, which would
have been (mildly) selected against in the \coshalos\ sample, may
affect its CGM. However, it's unlikely we could distinguish these
effects from simple stochastic variations for a single galaxy.

The environment of M31 could plausibly play a role in shaping its
CGM. \cite{burchett2016} found a smaller detection rate (covering
factor) of \ion{C}{4} for galaxies in high density environments
(assessed over $\rho \approx 1.5$ Mpc). They do not find a
corresponding difference in \HI\ covering factor, but their dynamic
range in \NHI\ is small due to saturation of \lya\ (they cannot probe
columns higher than $\logNHI \sim 14.5$). However, the Local Group
would not be included in their high density category, as only 2-3
galaxies would make their luminosity cut ($M_r < -19$ mag) for
identifying galaxies in their density counting scheme. We note that
\citet{nuza2014} have presented simulations of a Local Group-like pair
of halos, both of which show high covering factors. They find near
unity (differential) covering factors for gas at $\logNHI \ga 17.85$
to 30-50 kpc in their two halos. However, this is based on only one
pair of halos seen at a single time, so we cannot yet draw strong
conclusions based on these simulations.

In general higher covering fractions are found from metal absorption
line searches. Several studies of strong \MgII\ absorption (typically
$W_r \ga 0.3$ m\AA ) have found quite high covering factors (e.g.,
$>0.6$) within $\sim\Rvir$ \citep{chen2010, nielsen2013}, even for
relatively massive galaxies like M31. Some of this difference has to
do with the broad range of \HI\ column densities probed by \MgII\
absorption, which can trace gas over nearly 5 orders of magnitude in
\NHI\ \citep[gas with $\log N(\mbox{\MgII}) \sim 13$ --
$W_r(2796) \sim 0.3$ m\AA\ -- can probe solar metallicity gas with
$\logNHI \sim 16$ or damped Ly$\alpha$ systems with $<1/300$ of the
solar metallicity; ][]{wotta2016}.

We note that \cite{rao2013} searched for the signature of \MgII\ and
other low-ion absorption from the halo of M31 using low-resolution COS
spectra. They probed impact parameters $13 \la \rho \la 112$ kpc with
10 AGNs, finding absorption in four sight lines, all of which reside
at $\rho \la 40$ kpc and within the $\logNHI \sim 18.3$ contours of
the 21-cm measurements. Their observations, sensitive to
$W_r(2796) \sim 0.3 - 0.5$ \AA, targeted AGNs projected along
M31's major axis. The lack of absorption in their four targets
at $40 \la \rho \la 112$ kpc is perhaps not surprising given the small
number of sight lines and the low covering factor of high column
density gas we find (their low-resolution observations have
sensitivity to only $W_r(2796) \sim 300$ m\AA, whereas Project AMIGA's
sensitivity is $\sim20$ m\AA).

Roughly half of the sight lines observed in this work have been
observed by \hst /COS at intermediate resolution (G130M+G160M) as part
of Project AMIGA. These observations, a subset of which were published
by \citetalias{lehner2015}, are sensitive to relatively weak metal
line absorption, notably in transitions from the ions \CII,
\ion{C}{4}, \SiII, \SiIII, \ion{Si}{4}. The preliminary data from this
program suggest a quite high metal line covering factor. One of the
initial goals for Project AMIGA GBT was to provide an \HI\ reference
for potential metallicity estimates, since \lya\ absorption from the
Andromeda galaxy is swamped by the Galactic absorption
trough. Unfortunately, with only upper limits on \NHI\ coupled with
potential beam-dilution effects, we cannot provide a hydrogen
reference for the metallicity determination. Without the hydrogen
reference, determining the metallicities along these sight lines will
not be possible \citep[on this point we disagree with the recent work
of ][]{koch2015}.

Better constraining the covering factor in the inner regions of M31's
CGM may be readily accomplished by simply observing more sight lines
(and more GBT observations are forthcoming). The \HI\ maps of
\citet{thilker2004} as analyzed by \citet{richter2012} provide some
guidance in this inner region, albeit at worse mass sensitivity than
the current observations and only at $\rho \la 50$ kpc. Pushing below
our current \HI\ column density limits, which are in the range of
$\logNHI \sim 17.4$ to 17.6 at $5\sigma$, is in principle
possible. However, in many of the cases presented here the limiting
factor in the \HI\ sensitivity is the quality of the spectral
baselines. The GBT baselines are already among the best available, but
we are working near the limit for the current instrumentation.

% \bull Stern model implications?

% In general the simulations of \citetalias{hafen2017} suggest galaxies
% like M31 contribute to the to the cosmological incidence of
% low-redshift LLSs \citep{ribaudo2011} roughly equally to other halos
% over the mass range $10 \la \log M_{\rm h} / \msun \la 12$, with M31 at the
% upper end of this range. While the covering factor of optically-thick
% gas about Andromeda-like halos is low as is their cosmological
% density, their large virial radii compensate for these factors.

% {\em Notes on the ionization fraction: } \citetalias{lehner2015} have
% noted that constraints can be placed on the ionization fraction of H
% based on a comparison of \ion{O}{1} absorption (which traces \HI) and
% \ion{Si}{2}+\ion{Si}{3}+\ion{Si}{4}, which traces all warm/warm-hot H
% (\HI\ + \ion{H}{2}). Assessment of the ionization fraction in this way
% indicates that the H is $\ga 97\%$ ionized. Thus, the ionization
% fraction of neutral hydrogen,
% $\log x({\rm H}^0) \equiv \log N({\rm H}^0)/N({\rm H}) \la -1.5$. This
% can be used to put constraints on the allowable metallicity as a
% function of...NHI, but not.

\section{Summary}
\label{sec:summary}

We have used the Green Bank Observatory's 100-m Robert C. Byrd Green
Bank Telescope to search for 21-cm emission from the CGM of our
neighbor M31. We detect no \HI\ emission, with column density limits
that overlap studies of QSO absorption line experiments (notably the
LLSs). Our principal conclusions are as follows.

\begin{enumerate}

\item We constrain the covering factor of optically-thick \HI\ with
  $\logNHI \ga 17.6$ about M31 to be
  $f_c (\le \Rvir) < \virialCovering$ (90\% confidence).

\item Our covering factor limits for these high \HI\ column densities
  are much lower than those found for metal lines about \lstar\
  galaxies, including M31 \citepalias{lehner2015}.

\item The covering factors derived here are also discrepant from
  recent measurements by the \coshalos\ project at the same \HI\
  column densities \citep{prochaska2017}. The origin of this
  difference is not clear. It may be related to characteristics of M31
  that are not represented in the \coshalos\ sample, although the
  difference could also be consistent with stochastic variations in
  M31's CGM with time.

\item The covering factors of optically-thick \HI\ about M31 are
  consistent those found in recent cosmological zoom simulations.

\end{enumerate}

\acknowledgements

Part of this manuscript was written at the 2016 Arthur M. Wolfe
Symposium in Astrophysics hosted by IMPS of UC Santa Cruz Department
of Astronomy. We thank the Esalen Institute for its great setting and
wonderful hospitality during that retreat. Support for HST Program
number 14268 was provided by NASA through a grant from the Space
Telescope Science Institute, which is operated by the Association of
Universities for Research in Astronomy, Incorporated, under NASA
contract NAS5-26555.  Some of this work was supported by NSF grants
AST-1212012 and AST-1517353 to Notre Dame and JHU. DJP recognizes
partial support from NSF CAREER grant AST-1149491.  Contributions by
ZH and CAFG were additionally supported through NSF grants AST-1412836
and AST-1517491 and through NASA grant NNX15AB22G.  This research made
use of Astropy, a community-developed core Python package for
Astronomy \citep[][]{robitaille2013}, and the matplotlib plotting
package \citep{hunter2007}.

\facility{GBT}

\software{Astropy \citep{robitaille2013},GBTIDL \citep{marganian2006},
  Matplotlib \citep{hunter2007}}

%%%%%%%%%%%%%%%%%%%%%%%%%%%%%%%%%%%%%%%%%%%%%%%%%%%%%%%%%%%%%%%%%%%%%%
\appendix
\section{Metal lines toward RBS~2055}
\label{sec:appendix}

\begin{figure}
  % % run pyplot_COSspectra.py
\epsscale{0.6}  

\plotone{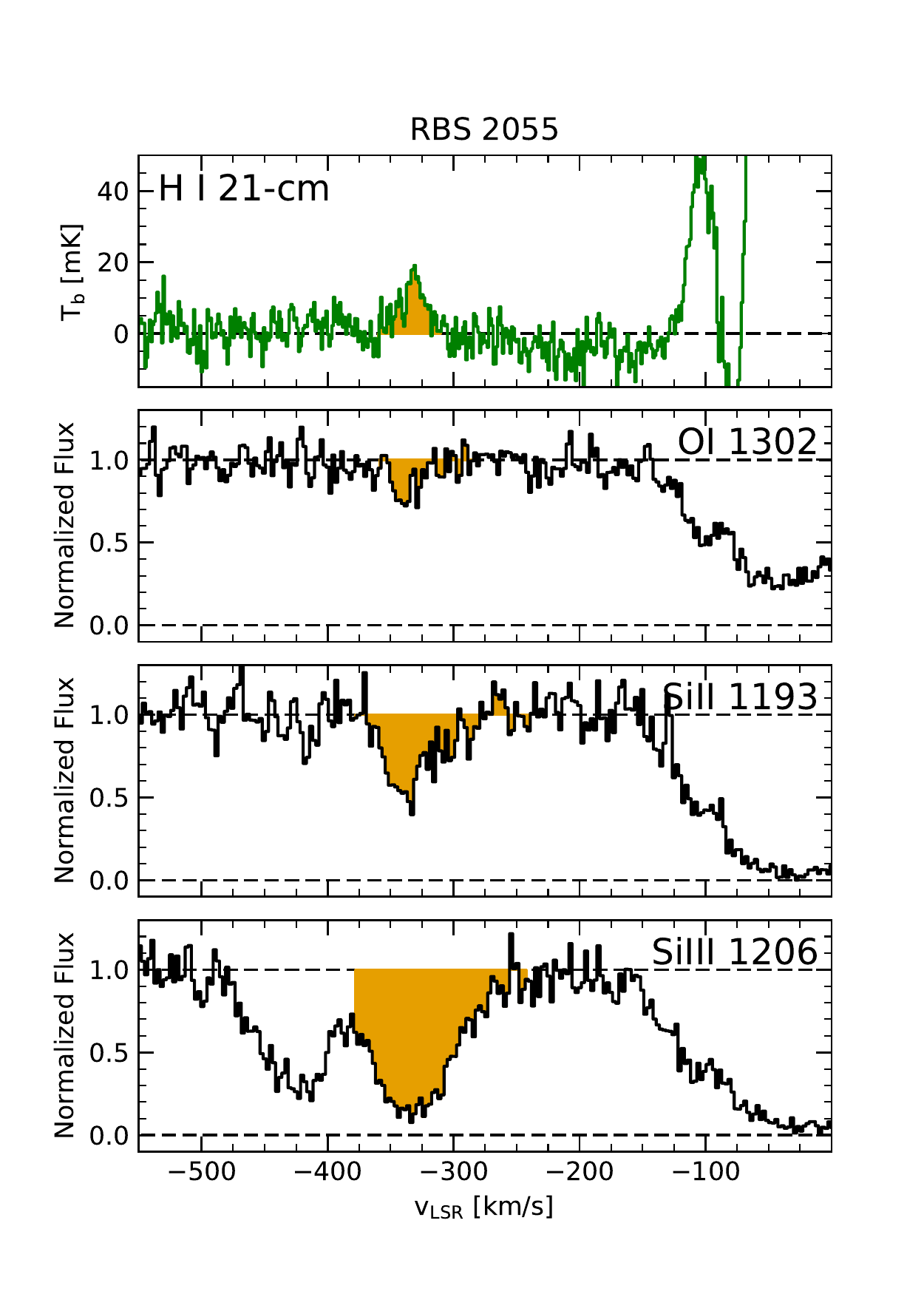}
\caption{COS absorption line spectra of the sight line to RBS~2055
  compared with the GBT \HI\ emission along this sight line (top). The
  metal absorption lines are \OI\ $\lambda1302$, \SiII\ $\lambda1193$,
  and \SiIII\ $\lambda1206$ (top to bottom). The shaded areas
  represent the region over which the column densities were determined
  (the GBT columns were determined via a Gaussian fit, the COS
  absorption columns via direct integration). The \SiIII\ absorption
  contains unresolved saturation, giving only a lower limit to the
  column density \citepalias[see][]{lehner2015}. In addition, the
  lesser saturation of the gas at $\vlsr \ga -310$ \kms\ relative to
  the principal component centered at $\vlsr \approx -335$ \kms\
  causes the central velocity of the much stronger \SiIII\ line to
  appear at higher velocities relative to the other transitions.
  These measurements together provide a measure of the metallicity,
  $[{\rm O/H}] = \MSmetallicity$. \label{fig:COSspectra}}
\end{figure}

The sight lines with detected \HI\ emission -- toward RBS~2055 and
RBS~2061 -- both lie close to the MS, and the velocity of this
emission is similar to expectations for the MS in these
directions. Figure \ref{fig:COSspectra} shows a few transitions from
our Project AMIGA HST/COS spectra of the sight line to RBS~2055
(N. Lehner et al., in prep.). We detect absorption from a wide range
of ions, including \CII, \OI, \SiII, \SiIII, and \ion{Si}{4}. These
are centered at $\vlsr \approx -335$ \kms, very slightly offset from
the centroid of our apparent \HI\ emission in this direction, but well
within the COS wavelength calibration uncertainties. The detection of
\OI\ is noteworthy, in that it implies a high column density of \HI,
in agreement with our direct 21-cm observations, and because \OI\ is
not detected along any of the other sight lines at large impact
parameters in our Project AMIGA data outside of the region projected
near to the MS.

%%%%%%%%%%%%%%%%%%%%%%%%%%%%%%%%%%%%%%%%%%%%%%%%%%%%%%%%%%%%%%%%%%%%%%
%%%%%%%%%%%%%%%%%%%%%%%%%%%%%%%%%%%%%%%%%%%%%%%%%%%%%%%%%%%%%%%%%%%%%%

The detection of \OI\ is also significant because the ionization
fractions of \OI\ and \HI\ are coupled by a strong charge exchange
reaction \citep[e.g.,]{chambaud1980, stancil1999}. Thus ionization
corrections that might apply in transforming measures of
$N(\mbox{\OI})/\NHI$ into the O/H abundance ratio are generally very
small \citep[e.g.,][]{fox2013}. We measure an \OI\ column density in
this gas $\log N(\mbox{\OI}) = 13.62\pm0.11$. Comparing this directly
to our \HI\ column density measurements from Table
\ref{tab:MSdetections} (assuming ionization corrections are very
small) yields an estimated abundance $[{\rm O/H}] = \MSmetallicity$
(statistical errors only). This value is in good agreement with
several measurements made along the main body of the MS by Fox and
collaborators. \citet{fox2010,fox2013} have derived
$[{\rm O/H}] = -1.0$ to $-1.2$ in several regions along the MS. These
include a measurement along the sight line to NGC~7469, only
$\sim21^\circ$ from RBS~2055, which shows
$[{\rm O/H}] = -1.00 \pm 0.05$ (statistical errors only) for a column
density of $\logNHI = 18.62 \pm 0.03$ from GBT+Effelsberg observations
\citep{fox2013}. These measurements were made in the same way as our
abundance estimate: comparing metal lines (\OI\ in most cases) with
ground-based \HI\ 21-cm measurements. Thus, our metallicities should
be directly comparable to those of \citeauthor{fox2013}.

Given the location of these sight lines close to the great circle on
which the MS lies, as well as the overall agreement in the expected
velocities and metallicities, we conclude that the emission detected
toward RBS~2055 and the nearby RBS~2061 arises in the MS. There are
sight lines in our sample near to the MS that lack emission at these
velocities. However, this is not unexpected, as the MS is quite patchy
in this direction \citep[this region is referred to as the ``tip'' of
the MS by][]{nidever2010}. We also note that at velocities outside
of $-350 \la \vlsr \la -300$ \kms, the RBS~2055 and RBS~2061 sight
lines show no other emission. This is in contrast to the COS data
toward RBS~2055, which show well-detected absorption at
$\vlsr \sim-420$ \kms\ in ions such as \SiII, \SiIII, and \ion{C}{4},
presumably from the CGM of M31 (Figure \ref{fig:COSspectra} and
N. Lehner et al., 2017, in prep.).

Having concluded that the gas at $\vlsr \approx -330$ \kms\ in this
direction probes the MS, our absorption line data provide measures of
the physical conditions in the MS gas. Our observations include \SiII,
\SiIII, and \ion{Si}{4}. The summed column density of these ions is
$\log N({\rm Si}) \ge 13.73\pm0.02$. These ions together probe the
low- to moderate-ionization gas in the MS, including photoionized gas
and collisionally-ionized gas at $T \la 70,000$ K. \OI\ is a tracer of
the neutral gas content. Thus, the ratio of \OI\ to the total Si
column provides a measure of the ionization fraction of the gas
\citep[see ][]{lehner2015}. The \HI\ ionization fraction derived for
the cool/warm gas along this sightline is
$x({\rm H}^0) \equiv N({\rm H}^0)/N({\rm H}) > 0.97$, implying a total
column density of $\log N(\mbox{\HI + \ion{H}{2}}) > 19.6$, similar to
that found in a much larger sample of MS sight lines by
\citet{fox2014}. If one excludes the gas associated with \ion{Si}{4},
imagining it resides in much different physical conditions than the
lower ionization gas, the result is not much changed
($x({\rm H}^0) > 0.96$), since \ion{Si}{4} is only $\sim25\%$ of the
total column.

\bibliography{howk_master}

%%%%%%%%%%%%%%%%%%%%%%%%%%%%%%%%%%%%%%%%%%%%%%%%%%%%%%%%%%%%%%%%%%%%%%
%%%%%%%%%%%%%%%%%%%%%%%%%%%%%%%%%%%%%%%%%%%%%%%%%%%%%%%%%%%%%%%%%%%%%%
%% Tables
%%%%%%%%%%%%%%%%%%%%%%%%%%%%%%%%%%%%%%%%%%%%%%%%%%%%%%%%%%%%%%%%%%%%%%
\pagebreak
\clearpage

%%%%%%%%%%%%%%%%%%%%%%%%%%%%%%%%%%%%%%%%%%%%%%%%%%%%%%%%%%%%%%%%%%%%%%
%% Table 1: pointings and sensitivities
%%
%% Table 1 is produced by GBT-AMIGA/Data-Reduced/make_tablePointings.pro
%%

\begin{deluxetable}{lcccccrc}
\tablenum{1}
\tablewidth{0pc}
\tablecaption{GBT Targets near M31 \label{tab:targets}}
\tabletypesize{\footnotesize}
\tablehead{\colhead{QSO name} & \colhead{RA} &  \colhead{Dec} &
   \colhead{$\Delta \theta$\tablenotemark{a}} & \colhead{$\rho$\tablenotemark{b}} &   \colhead{$\rho/R_{vir}$\tablenotemark{c}} &   \colhead{$\sigma_{\rm b}$\tablenotemark{d}} &   \colhead{$\log N({\rm H\, I})$\tablenotemark{e}} \\
\colhead{} & \colhead{(J2000)} &  \colhead{(J2000)} & \colhead{$(\degr)$} &
\colhead{(kpc)} &   \colhead{} &   \colhead{(mK)} &   \colhead{(5$\sigma$ sens.)}}
\startdata
RXJ0048.3+3941          & $00:48:19$ & $+39:41:11.0$ & $ 1.9$ & $ 24$ & $ 0.08$ & $ 2.8$ & $17.43$ \\
HS0033+4300             & $00:36:23$ & $+43:16:40.2$ & $ 2.3$ & $ 30$ & $ 0.10$ & $10.1$ & $17.55$ \\
HS0058+4213             & $01:01:31$ & $+42:29:35.6$ & $ 3.7$ & $ 48$ & $ 0.16$ & $10.4$ & $17.56$ \\
RXSJ0043.6+372521       & $00:43:42$ & $+37:25:19.0$ & $ 3.9$ & $ 50$ & $ 0.17$ & $ 9.2$ & $17.50$ \\
ZW535.012               & $00:36:20$ & $+45:39:53.2$ & $ 4.5$ & $ 59$ & $ 0.20$ & $ 9.6$ & $17.52$ \\
Q0030+3700              & $00:30:17$ & $+37:00:54.0$ & $ 4.9$ & $ 64$ & $ 0.21$ & $ 9.4$ & $17.51$ \\
BLANK                   & $01:08:27$ & $+38:58:32.0$ & $ 5.4$ & $ 71$ & $ 0.24$ & $ 7.4$ & $17.41$ \\
MS0108.4+3859           & $01:11:16$ & $+39:15:00.0$ & $ 5.8$ & $ 76$ & $ 0.25$ & $ 1.9$ & $17.27$ \\
RXSJ005050.6+353645     & $00:50:50$ & $+35:36:43.0$ & $ 5.9$ & $ 76$ & $ 0.26$ & $ 6.3$ & $17.34$ \\
2E0111.0+3851           & $01:13:54$ & $+39:07:44.0$ & $ 6.3$ & $ 82$ & $ 0.28$ & $ 7.9$ & $17.44$ \\
IRAS00040+4325          & $00:06:36$ & $+43:42:29.0$ & $ 7.1$ & $ 92$ & $ 0.31$ & $ 9.7$ & $17.53$ \\
RXSJ011848.2+383626     & $01:18:49$ & $+38:36:19.0$ & $ 7.4$ & $ 96$ & $ 0.32$ & $ 9.8$ & $17.53$ \\
RXJ0117.7+3637          & $01:17:45$ & $+36:37:14.9$ & $ 8.2$ & $107$ & $ 0.36$ & $ 6.8$ & $17.37$ \\
SDSSJ001847.44+341209.5 & $00:18:47$ & $+34:12:09.6$ & $ 8.5$ & $111$ & $ 0.37$ & $ 9.9$ & $17.54$ \\
MRK352                  & $00:59:53$ & $+31:49:37.1$ & $10.0$ & $131$ & $ 0.44$ & $ 9.9$ & $17.54$ \\
RXJ0028.1+3103          & $00:28:10$ & $+31:03:48.1$ & $10.6$ & $138$ & $ 0.46$ & $ 8.7$ & $17.48$ \\
NGC513                  & $01:24:26$ & $+33:47:57.9$ & $11.1$ & $145$ & $ 0.48$ & $ 7.0$ & $17.39$ \\
KAZ238                  & $00:00:58$ & $+33:20:38.5$ & $11.5$ & $149$ & $ 0.50$ & $ 9.0$ & $17.50$ \\
MRK1158                 & $01:34:59$ & $+35:02:22.1$ & $12.0$ & $156$ & $ 0.52$ & $ 7.7$ & $17.43$ \\
2MASSJ00413+2816        & $00:41:18$ & $+28:16:40.8$ & $13.0$ & $169$ & $ 0.56$ & $10.6$ & $17.57$ \\
FBS0150+396             & $01:53:06$ & $+39:55:45.2$ & $13.4$ & $174$ & $ 0.58$ & $ 9.3$ & $17.51$ \\
3C48                    & $01:37:41$ & $+33:09:35.1$ & $13.6$ & $176$ & $ 0.59$ & $14.8$ & $17.71$ \\
4C25.01                 & $00:19:39$ & $+26:02:52.3$ & $16.0$ & $206$ & $ 0.69$ & $ 7.0$ & $17.39$ \\
PG0052+251              & $00:54:52$ & $+25:25:38.9$ & $16.0$ & $207$ & $ 0.69$ & $ 6.9$ & $17.38$ \\
2MASSJ00294+2424        & $00:29:24$ & $+24:24:29.0$ & $17.1$ & $220$ & $ 0.74$ & $10.7$ & $17.57$ \\
RXSJ015536.7+311525     & $01:55:36$ & $+31:15:17.0$ & $17.7$ & $228$ & $ 0.76$ & $ 6.8$ & $17.37$ \\
RBS2055                 & $23:51:52$ & $+26:19:32.5$ & $18.3$ & $235$ & $ 0.79$ & $ 6.1$ & $17.33$ \\
RBS2061                 & $23:55:48$ & $+25:30:31.6$ & $18.5$ & $238$ & $ 0.80$ & $ 6.0$ & $17.32$ \\
3C66A                   & $02:22:39$ & $+43:02:07.8$ & $18.5$ & $239$ & $ 0.80$ & $12.6$ & $17.64$ \\
RXJ0053.7+2232          & $00:53:46$ & $+22:32:22.1$ & $18.9$ & $243$ & $ 0.81$ & $ 6.6$ & $17.36$ \\
MRK930                  & $23:31:58$ & $+28:56:49.9$ & $18.9$ & $244$ & $ 0.81$ & $ 5.6$ & $17.29$ \\
RXJ0048.7+2127          & $00:48:45$ & $+21:27:15.9$ & $19.9$ & $255$ & $ 0.85$ & $12.0$ & $17.62$ \\
MRK357                  & $01:22:40$ & $+23:10:14.7$ & $19.9$ & $256$ & $ 0.85$ & $ 8.2$ & $17.46$ \\
3C59                    & $02:07:02$ & $+29:30:45.9$ & $20.7$ & $266$ & $ 0.89$ & $11.9$ & $17.62$ \\
PG0117+213              & $01:20:17$ & $+21:33:46.2$ & $21.2$ & $272$ & $ 0.91$ & $ 6.3$ & $17.34$ \\
HS0137+2329             & $01:40:35$ & $+23:44:51.0$ & $21.3$ & $272$ & $ 0.91$ & $ 6.7$ & $17.37$ \\
KUV02196+3253           & $02:22:31$ & $+33:06:21.2$ & $21.4$ & $274$ & $ 0.91$ & $ 6.7$ & $17.37$ \\
RXJ0029.0+1957          & $00:29:03$ & $+19:57:10.0$ & $21.5$ & $275$ & $ 0.92$ & $ 7.9$ & $17.44$ \\
RXJ0044.9+1921          & $00:44:59$ & $+19:21:41.0$ & $21.9$ & $280$ & $ 0.94$ & $ 9.9$ & $17.54$ \\
MRK335                  & $00:06:19$ & $+20:12:10.5$ & $22.4$ & $287$ & $ 0.96$ & $ 9.6$ & $17.52$ \\
UGC1098                 & $01:32:16$ & $+21:24:38.9$ & $22.4$ & $287$ & $ 0.96$ & $ 9.4$ & $17.52$ \\
RXSJ023231.4+340435     & $02:32:33$ & $+34:04:27.9$ & $22.8$ & $291$ & $ 0.97$ & $ 9.7$ & $17.53$ \\
RXSJ225148.5+341937     & $22:51:47$ & $+34:19:28.9$ & $22.9$ & $292$ & $ 0.97$ & $10.1$ & $17.55$ \\
MRK1148                 & $00:51:54$ & $+17:25:58.4$ & $23.9$ & $304$ & $ 1.02$ & $ 6.4$ & $17.35$ \\
RBS2005                 & $23:25:54$ & $+21:53:14.0$ & $25.2$ & $320$ & $ 1.07$ & $ 8.0$ & $17.44$ \\
MRK1179                 & $02:33:22$ & $+27:56:13.1$ & $26.2$ & $332$ & $ 1.11$ & $ 9.2$ & $17.51$ \\
PG0003+158              & $00:05:59$ & $+16:09:49.0$ & $26.3$ & $333$ & $ 1.11$ & $ 9.6$ & $17.52$ \\
BLANK                   & $22:42:39$ & $+29:43:31.5$ & $26.8$ & $339$ & $ 1.13$ & $ 9.1$ & $17.50$ \\
\enddata
\tablenotetext{a}{Anglular separation from the center of M31.}
\tablenotetext{b}{We adopt 752 kpc as the distance of M31 from the sun \citep{riess2012}. }
\tablenotetext{c}{We assume $\Rvir = 300$ kpc following \citet{lehner2015}.}
\tablenotetext{d}{RMS brightness temperature noise per 0.6 \kms\ channel.}
\tablenotetext{e}{The 5$\sigma$ column density sensitivities (in \column) for each sight line. Any detections are summarized in Table \ref{tab:MSdetections}.}
\end{deluxetable}

%%%%%%%%%%%%%%%%%%%%%%%%%%%%%%%%%%%%%%%%%%%%%%%%%%%%%%%%%%%%%%%%%%%%%%
%% Table 2: H I detections
%%
%% Table 2 is produced by GBT-AMIGA/Data-Reduced/pymake_tableDetections.py
%%

\begin{deluxetable}{lcccc}
\tablenum{2}
\tablecaption{\HI\ Detections\label{tab:MSdetections}}
\tablehead{\colhead{QSO Name} & \colhead{$\rho$} & \colhead{$\log N(\mbox{\HI})$} & \colhead{$\langle \vlsr \rangle$} & \colhead{FWHM}\\ \colhead{ } & \colhead{(kpc)} & \colhead{[cm$^{-2}$]} & \colhead{(\kms)} & \colhead{(\kms)}}
\startdata
RBS2055 & 235 & $17.88\pm0.05$ & $-331.5\pm1.0$ & $25.1\pm2.3$ \\
RBS2061 & 238 & $17.79\pm0.08$ & $-334.0\pm1.7$ & $29.1\pm4.0$
\enddata
\end{deluxetable}

%%%%%%%%%%%%%%%%%%%%%%%%%%%%%%%%%%%%%%%%%%%%%%%%%%%%%%%%%%%%%%%%%%%%%%
%% Table 3: covering factors
%%
%% Table 3 is produced by GBT-AMIGA/Data-Reduced/pymake_coveringfactors.py
%%

\begin{deluxetable}{cccc}
\tablenum{3}
\tablecaption{\HI\ Covering Factors\label{tab:coveringfactor}}
\tablehead{\colhead{$\rho$ (kpc)} & \colhead{$\rho / R_{\rm vir}$} & \colhead{$f_c(\le\rho)$\tablenotemark{a}} & \colhead{$N$\tablenotemark{b}}}
\startdata
50 & 0.17 & $< 0.369$ & 4 \\
75 & 0.25 & $< 0.250$ & 7 \\
100 & 0.33 & $< 0.162$ & 12 \\
125 & 0.42 & $< 0.142$ & 14 \\
150 & 0.50 & $< 0.114$ & 18 \\
175 & 0.58 & $< 0.099$ & 21 \\
200 & 0.67 & $< 0.095$ & 22 \\
225 & 0.75 & $< 0.085$ & 25 \\
250 & 0.83 & $< 0.069$ & 31 \\
275 & 0.92 & $< 0.057$ & 38 \\
300 & 1.00 & $< 0.051$ & 43 \\
325 & 1.08 & $< 0.049$ & 45 \\
350 & 1.17 & $< 0.046$ & 48
\enddata
\tablenotetext{a}{Upper limits are at 90\%\ confidence.} \tablenotetext{b}{Number of sight lines considered in covering factor determination.}
\end{deluxetable}

%%%%%%%%%%%%%%%%%%%%%%%%%%%%%%%%%%%%%%%%%%%%%%%%%%%%%%%%%%%%%%%%%%%%%%
%% Table 4: covering factors with MS detections
%%
%% Table 4 is produced by GBT-AMIGA/Data-Reduced/pymake_coveringfactors.py
%%

\begin{deluxetable}{cccccc}
\tablenum{4}
\tablecaption{\HI\ Covering Factors with MS Detections\label{tab:MScoveringfactor}}
\tablehead{\colhead{$\rho$ (kpc)} & \colhead{$\rho / R_{\rm vir}$} & \colhead{$f_c(\le \rho)_{0.1}$\tablenotemark{a}} & \colhead{$f_c(\le \rho)_{0.5}$\tablenotemark{a}} & \colhead{$f_c(\le \rho)_{0.9}$\tablenotemark{a}} & \colhead{$N$\tablenotemark{b}}}
\startdata
250 & 0.83 & 0.035 & 0.083 & 0.158 & 31 \\
275 & 0.92 & 0.029 & 0.068 & 0.131 & 38 \\
300 & 1.00 & 0.025 & 0.060 & 0.116 & 43 \\
325 & 1.08 & 0.024 & 0.058 & 0.112 & 45 \\
350 & 1.17 & 0.023 & 0.054 & 0.105 & 48
\enddata
\tablenotetext{a}{The 10\%, 50\%\ (median), and 90\%\ quantiles for the \HI\ covering factors assuming the probable MS detections are associated with M~31.} \tablenotetext{b}{Number of sight lines considered in covering factor determination.}
\end{deluxetable}

\end{document}